\documentclass[12pt]{iopart}
\usepackage{graphicx}
\usepackage{iopams}
\usepackage{subfigure}
\begin{document}
\title{Lasing in metamaterial nanostructures}
\author{Anan Fang$^1$, Thomas Koschny$^{1,2}$,
Costas M Soukoulis$^{1,2}$}
\address{$^1$ Ames Laboratory and Department of Physics and Astronomy, Iowa State University, Ames, Iowa 50011, USA}
\address{$^2$ Institute of Electronic Structure and Laser, FORTH, and Department of Materials Science and Technology, University of Crete, 71110 Heraklion, Crete, Greece}
\ead{soukoulis@ameslab.gov}
\begin{abstract}
 A self-consistent computational scheme is presented for one dimensional (1D) and two dimensional (2D) metamaterial systems with gain incorporated into the nanostructures. The gain is described by a generic four-level system. The loss compensation and the lasing behavior of the metamaterial system with gain are studied. A critical pumping rate exists for compensating the losses of the metamaterial. There exists a wide range of input signals where the composite system behaves linearly. Nonlinearities arise for stronger signals due to gain depletion. The retrieved effective parameters are presented for one layer of gain embedded in two layers of Lorentz dielectric rods and split ring resonators with two different gain inclusions: (1) gain is embedded in the gaps only and (2) gain is surrounding the SRR. When the pumping rate increases, there is a critical pumping rate that the metamaterial system starts lasing. 
\end{abstract}
\submitto{\JOA}
\maketitle

\section{Introduction}

The field of metamaterials \cite {1,2} is driven by fascinating and far-reaching theoretical visions such as, e.g., perfect lenses \cite{3}, invisibility cloaking \cite{4,5}, and enhanced optical nonlinearities \cite{6}. This emerging field has seen spectacular experimental progress in recent years \cite{1,2}. Yet, losses are orders of magnitude too large for the envisioned applications. Achieving such reduction by further design optimization appears to be out of reach. Thus, incorporation of active media (gain) might reduce the losses. The procedure would be to simply inject an electrical current into the active medium, leading to gain and hence to compensation of the losses. However, experiments on such intricate active nanostructures do need guidance by theory via self-consistent calculations (using the semi-classical theory of lasing) for realistic gain materials that can be incorporated into or close to dispersive media to reduce the losses at THz or optical frequencies. The need for \textit{self-consistent} calculations stems from the fact that increasing the gain in the metamaterial, the metamaterial properties change, in turn changes the coupling to the gain medium until a steady-state is reached. A specific geometry to overcome the severe loss problem of optical metamaterials and to enable bulk metamaterials with negative magnetic and electric response and controllable dispersion at optical frequencies is to interleave active, optically pumped gain material layers with the passive metamaterial lattice.

For reference, the best fabricated negative-index material operating at around $1.4\,\mathrm{\mu m}$ wavelength \cite{7} has shown a figure of merit, $\mathrm{FOM} = -\mathrm{Re}(n)/\mathrm{Im}(n) \approx 3$, where $n$ is the effective refractive index.  This experimental result is equivalent to an absolute absorption coefficient of $\alpha=3\times10^4 \,\mathrm{cm}^{-1}$, which is even larger than the absorption of typical direct-gap semiconductors such as, e.g., GaAs (where $\alpha=10^4 \,\mathrm{cm}^{-1}$). So it looks difficult to compensate the losses with this simple type of analysis, which assumes that the bulk gain coefficient is needed. However, the effective gain coefficient, derived from self-consistent microscopic calculations, is a more appropriate measure of the combined system of metamaterial and gain.  Due to pronounced local-field enhancement effects in the spatial vicinity of the dispersive metamaterial, the effective gain coefficient can be substantially larger than its bulk counterpart. While early models \cite {8,9,10,11,12} using simplified gain-mechanisms such as explicitly forcing negative imaginary parts of the local gain material's response function produce unrealistic strictly linear gain, our self-consistent approach presented below allows for determining the range of parameters for which one can realistically expect linear amplification and linear loss compensation in the metamaterial \cite {13}. To fully understand the coupled metamaterial-gain system, we have to deal with time-dependent wave equations in metamaterial systems by coupling Maxwell's equations with the rate equations of electron populations describing a multi-level gain system in semi-classical theory \cite{14}.

This paper aims to apply a detailed computational model to the problem of metamaterials with gain. In \sref {theory}, we present the semi-classical theory of lasing and describe in detail the computational approach. In \sref {examples}, we verify that our code agrees well with simple soluble models (gain material only). In addition, our code is applied to 1D superlattice of gain and negative index layers. Next, a 2D problem is considered, which is a square lattice of Lorentz dielectric cylinders with layers of gain material. Finally, a 2D split ring resonator (SRR) with gain material inclusions is considered. Gain can compensate the losses and lasing (spasing) is achieved in our numerical simulations. In \sref {conclusion}, we present our conclusions.

\section{Theoretical and numerical model} \label {theory}
\begin{figure}
\centering
  \includegraphics[angle=-90, width=0.4\textwidth]{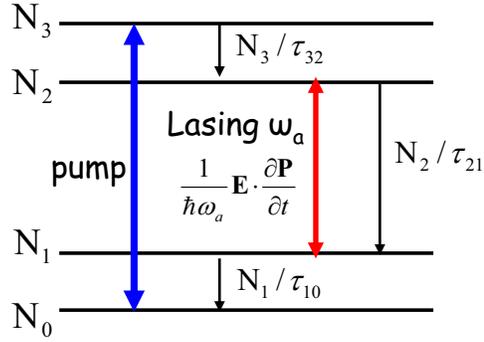}
\caption {%
 (Color online)
 Schematic of the four-level atomic system model.}
\label{fig1}
\end{figure}

The gain atoms are embedded in host medium and described by a generic four-level system, as shown in \fref {fig1}. All quantities including the fields and occupation numbers are tracked at each point in space and take into account energy exchange between gain atoms and fields, external pumping and non-radiative decays \cite {14}. Electrons are pumped by an external mechanism from the ground state level ($N_0$) to the third level ($N_3$). After a short lifetime $\tau_{32}$, they quickly relax into the metastable second level ($N_2$). The second level ($N_2$) and the first level ($N_1$) are called as the upper and lower lasing levels, respectively. Electrons  can transfer both radiatively (spontaneous and stimulated emissions) and non-radiatively from the upper to the lower lasing level. At last, they transfer quickly and non-radiatively from the first level ($N_1$) to the ground state level ($N_0$). The energies of ground state and the third level are $E_0$ and $E_3$. In optical pumping mechanism, electrons are raised from the ground state level ($N_0$) to the third level ($N_3$) by an external electromagnetic wave with the pumping frequency $\omega_b = (E_3-E_0)/\hbar$, which is chosen to be  $4\pi \times 10^{14}\,\mathrm {Hz}$ in our simulations. The local intensity of the pumping EM wave varies with the position and determines the pumping rate at each point. The lifetimes and energies of the upper and lower lasing levels are $\tau_{21} ,\ E_2$ and $\tau_{10} ,\ E_1$, respectively. The center frequency of the radiation is $\omega_a=(E_2-E_1)/\hbar$, which is chosen to be $2\pi \times 10^{14}\,\mathrm {Hz}$. The parameters $\tau_{32}$, $\tau_{21}$ and $\tau_{10}$ are chosen $5\times 10^{-14}\, \mathrm {s}$, $5\times 10^{-12}\, \mathrm {s}$ and $5\times 10^{-14}\, \mathrm {s}$, respectively. The total electron density, $N_0(t=0) = N_0(t)+N_1(t)+N_2(t)+N_3(t)=5.0\times 10^{23}\, \mathrm {/m^3}$.

The time-dependent Maxwell equations are given by

\begin{equation}
\label {MaxwellEquations}
\eqalign {
\nabla\times \mathbf {E}  =  -\partial \mathbf {B}/\partial t  \\
\nabla\times \mathbf {H}  =  \varepsilon \varepsilon_o \partial \mathbf {E}/\partial t + \partial \mathbf {P}/\partial t,
}
\end{equation}
where $\mathbf {B}=\mu\mu_o \mathbf {H}$ and $\mathbf {P}=\sum_{i=a, b} \mathbf {P}_i$ is the electric polarization density of the gain material. ($\mathbf {P}_a$ is the induced electric polarization density on the atomic transition between the upper ($N_2$) and lower ($N_1$) lasing levels, and $\mathbf {P}_b$ is between the ground state level ($N_0$) and the third level ($N_3$).) The induced electric polarizations behave as harmonic oscillators and couple to the local $\mathbf {E}$ field, which is propagated by Maxwell equations. The polarization density $\mathbf {P}_i(\mathbf {r},t)$ obeys locally the following equation of motion \cite {14}

\begin{equation}
\frac {\partial^2 \mathbf {P}_i(t)}{\partial t^2}+\Gamma_i \frac {\partial \mathbf {P}_i(t)}{\partial t}+\omega_i^2 \mathbf {P}_i(t)=-\sigma_i \Delta N_i(t) \mathbf {E}(t) \qquad (i = a, b),
\label {DrivenOscillator}
\end{equation}
where $\Gamma_i$ is the linewidth of the atomic transition $\omega_i$, $\sigma_i$ is the coupling strength of $\mathbf {P}_i$ to the electric field, and $\Delta N_a(\mathbf {r},t) = N_2(\mathbf {r},t)-N_1(\mathbf {r},t)$ and $\Delta N_b(\mathbf {r},t) = N_3(\mathbf {r},t)-N_0(\mathbf {r},t)$ are the population inversions that drive the polarizations. In our simulations, $\Gamma_a$ is chosen to be equal to $2\pi \times 5 \times 10^{12}\, \mathrm {Hz}$ or $2\pi \times 20 \times 10^{12}\, \mathrm {Hz}$ and  $\Gamma_b$ is equal to $2\pi \times 10 \times 10^{12}\, \mathrm {Hz}$. The values for $\sigma_a$ and $\sigma_b$ are taken to be $10^{-4}\,\mathrm{C^2/kg}$ and $5 \times 10^{-6}\,\mathrm{C^2/kg}$, respectively. From \eref {DrivenOscillator}, it can be easily derived \cite {14} that the atomic response of gain atoms has a Lorentzian lineshape and is homogeneously broadened. The occupation numbers at each spatial point vary according to the following rate equations,
\numparts
\begin{eqnarray}
\label {RateEquations_a}
  \frac {\partial N_3}{\partial t} = \frac {1}{\hbar \omega_b}\mathbf {E}\cdot \frac {\partial \mathbf {P}_b}{\partial t}-\frac {N_3}{\tau_{32}}, \\
\label {RateEquations_b}
  \frac {\partial N_2}{\partial t} = \frac{N_3}{\tau_{32}} +\frac{1}{\hbar\omega_a} \mathbf {E}\cdot\frac{\partial \mathbf {P}_a}{\partial t}-\frac{N_2}{\tau_{21}}, \\
\label {RateEquations_c}
  \frac {\partial N_1}{\partial t} = \frac{N_2}{\tau_{21}} - \frac{1}{\hbar\omega_a} \mathbf {E}\cdot\frac{\partial \mathbf {P}_a}{\partial t}  - \frac{N_1}{\tau_{10}},  \\
\label {RateEquations_d}
  \frac {\partial N_0}{\partial t} =-\frac {1}{\hbar \omega_b}\mathbf {E}\cdot \frac {\partial \mathbf {P}_b}{\partial t}+\frac {N_1}{\tau_{10}},
\end{eqnarray}
\endnumparts
where $\frac{1}{\hbar \omega_i}\mathbf {E} \cdot \frac {\partial \mathbf {P}_i}{\partial t}$ ($i = a, b$) is the induced radiation rate or excitation rate depending on its sign.

Instead of using an external EM wave to optically pump electrons from the ground state level ($N_0$) to the third level ($N_3$), we can simplify this process in \eref{RateEquations_a} and \eref{RateEquations_d}  by pumping electrons with a homogeneous pumping rate $\Gamma_{\mathrm {pump}}$, which is proportional to the optical pumping intensity in an experiment. This simplification is valid only if the gain slab is thin and the gain of the laser is low, because the real pumping rate depends on the local optical intensity and should be a function of position. We'll discuss this in more detail in \sref {GainOnly}. Based on this simplification, we can have the rate equations as follows,
\numparts
\begin{eqnarray}
\label {RateEquations1_a}
  \frac {\partial N_3}{\partial t} = \Gamma_{\mathrm {pump}}N_0-\frac {N_3}{\tau_{32}}, \\
\label {RateEquations1_b}
  \frac {\partial N_2}{\partial t} = \frac{N_3}{\tau_{32}} +\frac{1}{\hbar\omega_a} \mathbf {E}\cdot\frac{\partial \mathbf {P}_a}{\partial t}-\frac{N_2}{\tau_{21}}, \\
\label {RateEquations1_c}
  \frac {\partial N_1}{\partial t} = \frac{N_2}{\tau_{21}} - \frac{1}{\hbar\omega_a} \mathbf {E}\cdot\frac{\partial \mathbf {P}_a}{\partial t}  - \frac{N_1}{\tau_{10}},  \\
\label {RateEquations1_d}
  \frac {\partial N_0}{\partial t} =\frac {N_1}{\tau_{10}}-\Gamma_{\mathrm {pump}}N_0,
\end{eqnarray}
\endnumparts
Correspondingly, we only need to consider the electric polarization density $\mathbf {P}_a(\mathbf {r}, t)$ on the atomic transition between $N_2$ and $N_1$ in \eref {MaxwellEquations} and \eref {DrivenOscillator}.

In order to solve the behavior of the gain materials in the electromagnetic fields numerically, the finite-difference time-domain (FDTD) method is utilized \cite {15,16,17}. At the left and right ends of the computational space, perfect matched layers (PML) are used to impose the absorbing boundary condition (ABC). In the FDTD calculations, both the space and time are discretized into small steps compared to the characteristic space and time periods. In our simulations presented below, the discrete time and space steps are chosen to be $\Delta t = 1.67\times 10^{-17}\,\mathrm {s}$ and $\Delta x = 1.0\times 10^{-8}\,\mathrm {m}$ in \sref {GainOnly} and \sref {NIMInGain}, $\Delta t = 8.33\times 10^{-18}\,\mathrm {s}$ and $\Delta x = 5.0\times 10^{-9}\,\mathrm {m}$ in \sref {CylinderInGain}, and  $\Delta t = 8.33\times 10^{-19}\,\mathrm {s}$ and $\Delta x = 1.0\times 10^{-9}\,\mathrm {m}$ in \sref {SRRInGain}. The initial condition is that all electrons are in the ground state and all electric, magnetic and polarization fields are zero. Then the electrons are pumped from $N_0$ to $N_3$ optically or with a homogeneous pumping rate $\Gamma_{\mathrm {pump}}$. The system begins to evolve according to the equations above. 
\section{Examples  for metamaterials incorporated with gain}\label {examples}

\subsection{Gain material only}\label {GainOnly}

\begin{figure}
\center {
  \includegraphics[width=0.6\textwidth]{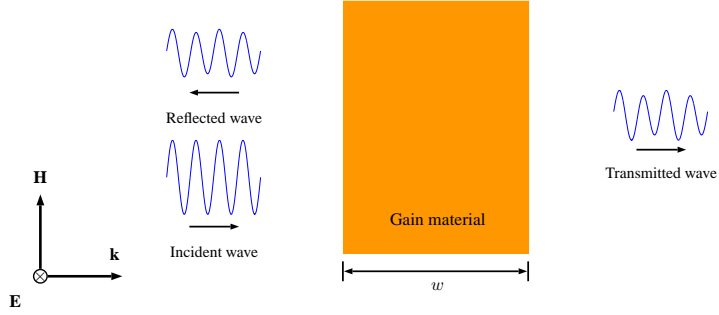}}
\caption {%
 (Color online)
 Schematic of gain material slab (shown in orange). The slab width $w$ takes different values in the cases we have examined.}
\label{fig2}
\end{figure}

\begin{figure*}
\centering
 \subfigure[]{
  \label {fig3a}
  \includegraphics[width=0.25\textwidth]{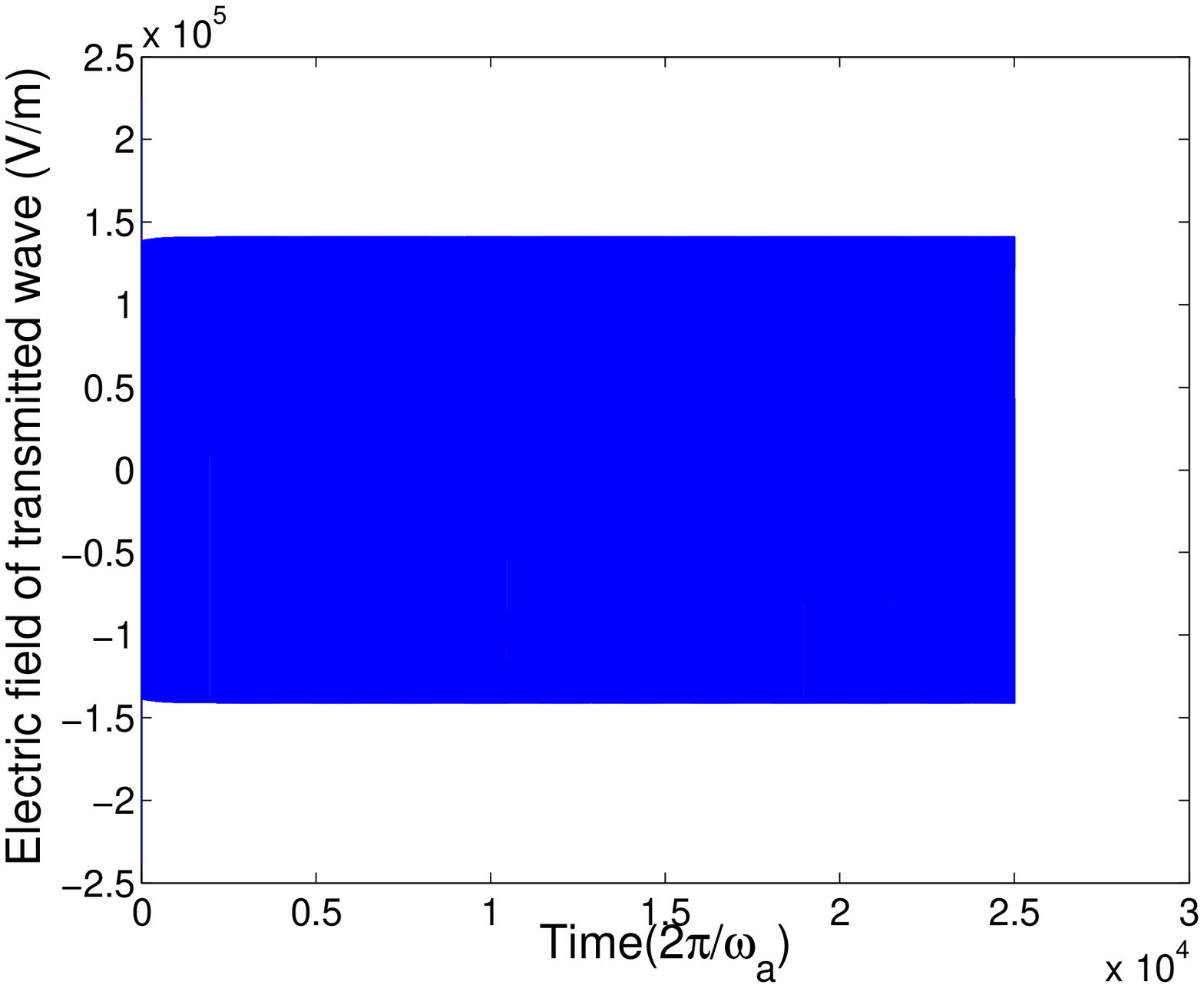}}
\qquad
\centering
 \subfigure[]{
  \label {fig3b}
  \includegraphics[width=0.25\textwidth]{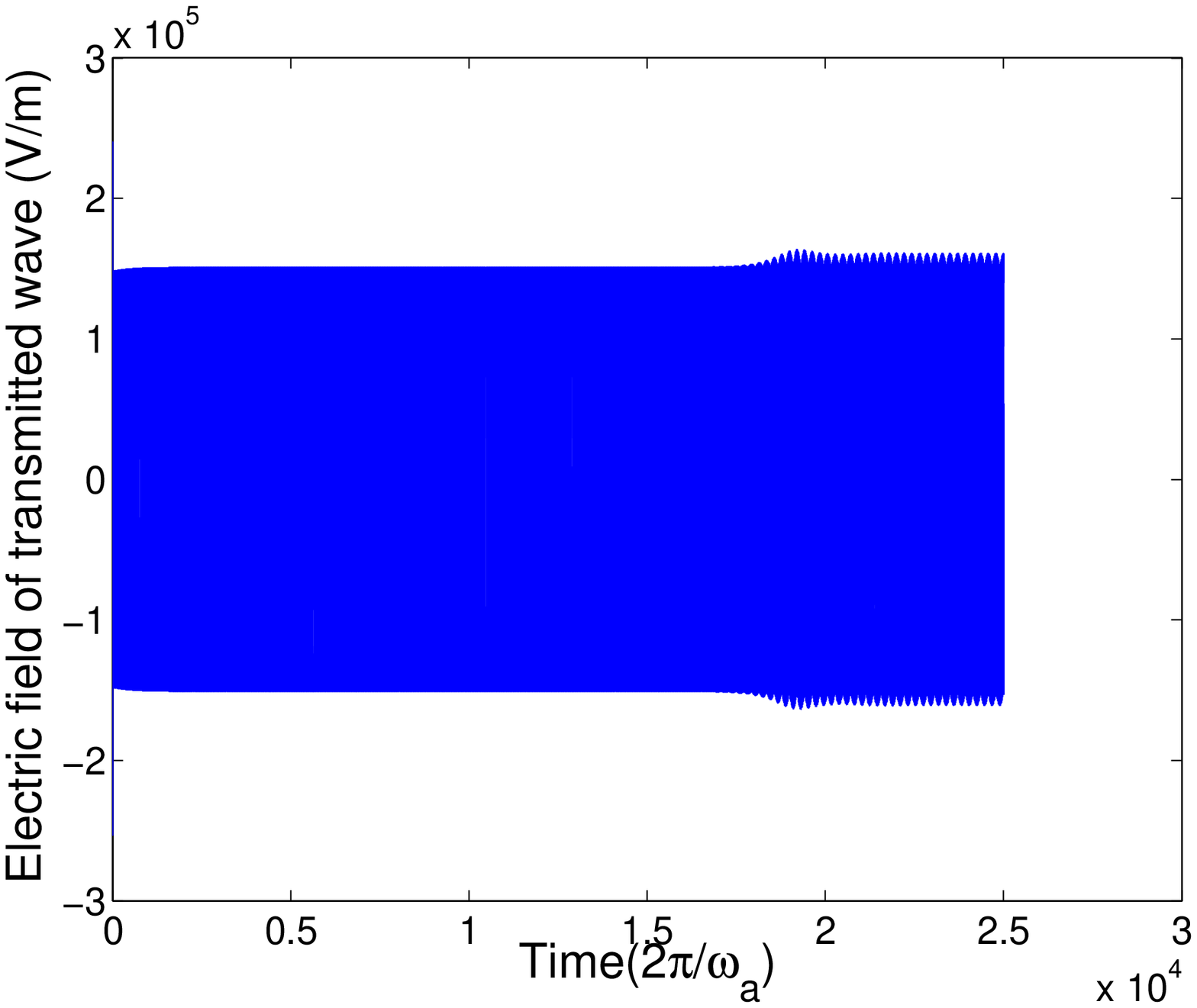}}
\qquad
\centering
 \subfigure[]{
  \label {fig3c}
  \includegraphics[width=0.25\textwidth]{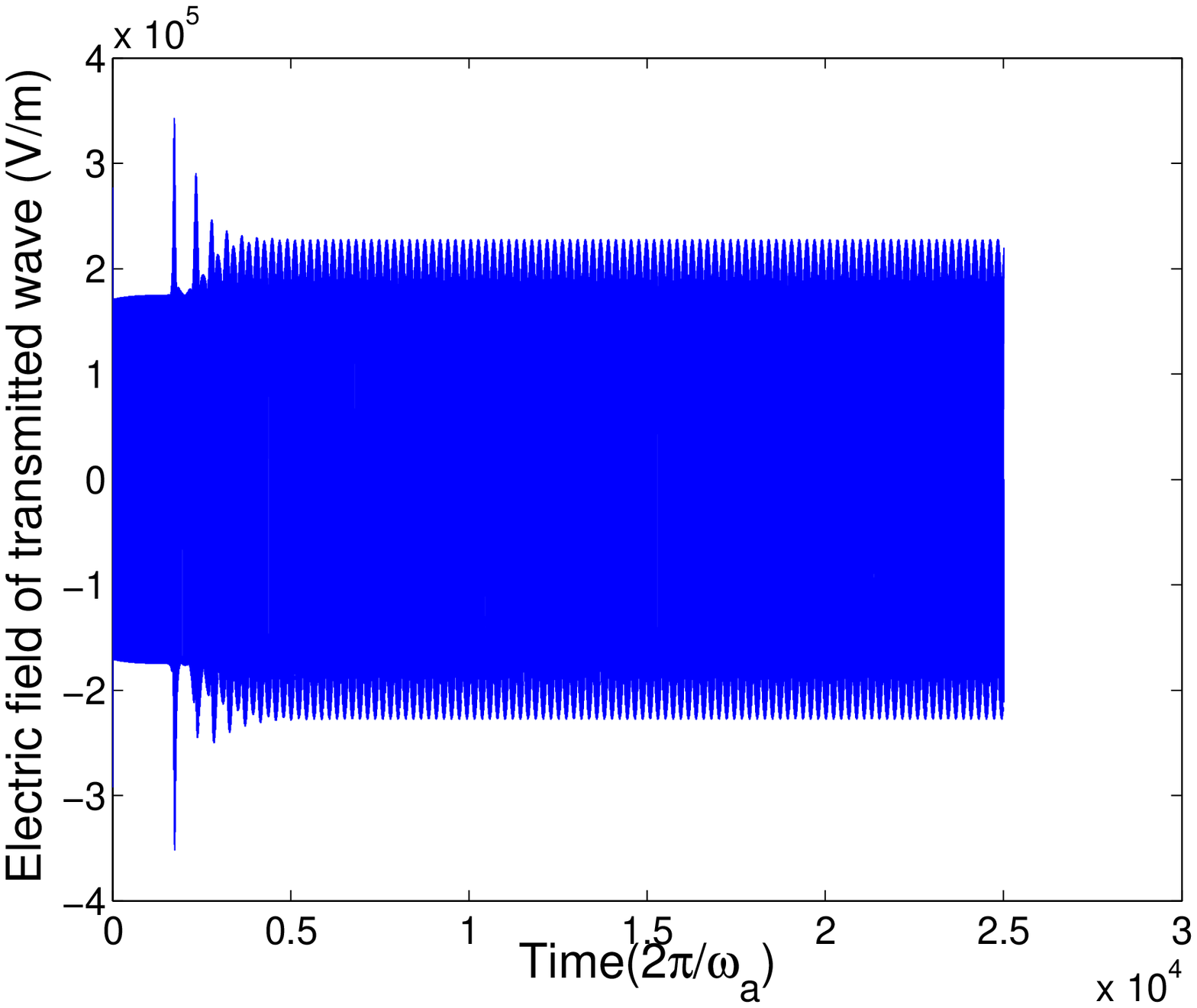}}
\\
\centering
 \subfigure[]{
  \label {fig3d}
  \includegraphics[width=0.25\textwidth]{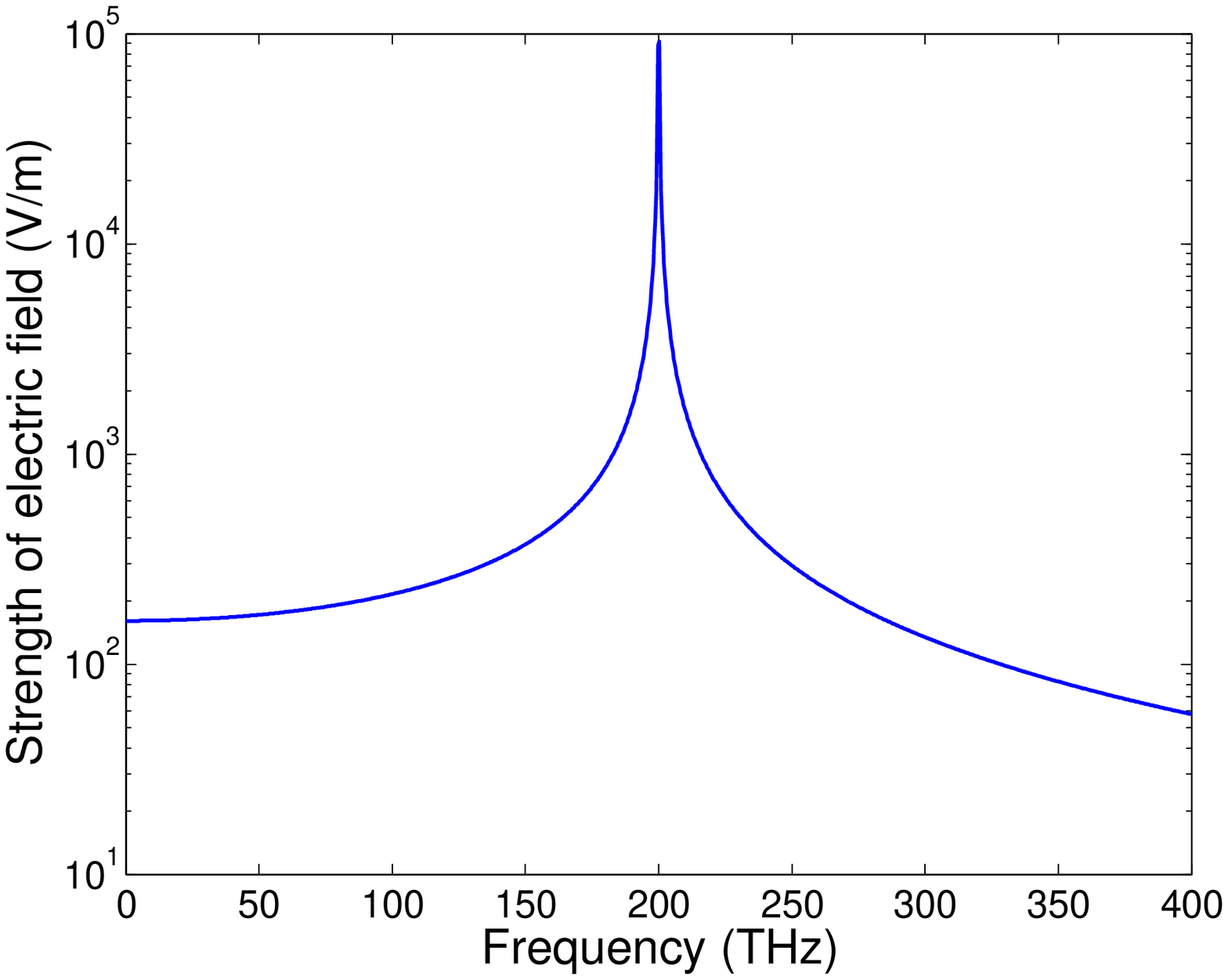}}
\qquad
\centering
 \subfigure[]{
  \label {fig3e}
  \includegraphics[width=0.25\textwidth]{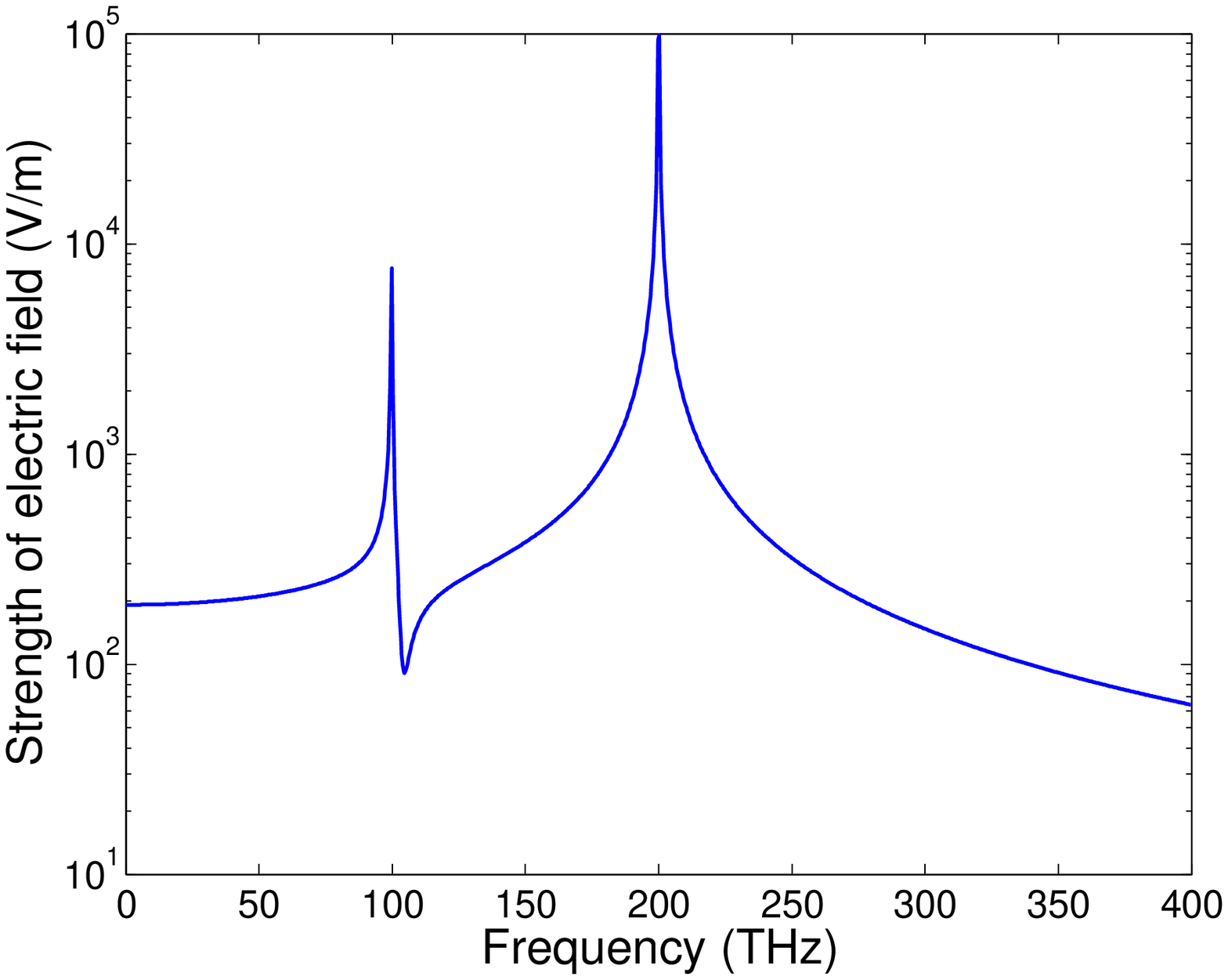}}
\qquad
\centering
 \subfigure[]{
  \label {fig3f}
  \includegraphics[width=0.25\textwidth]{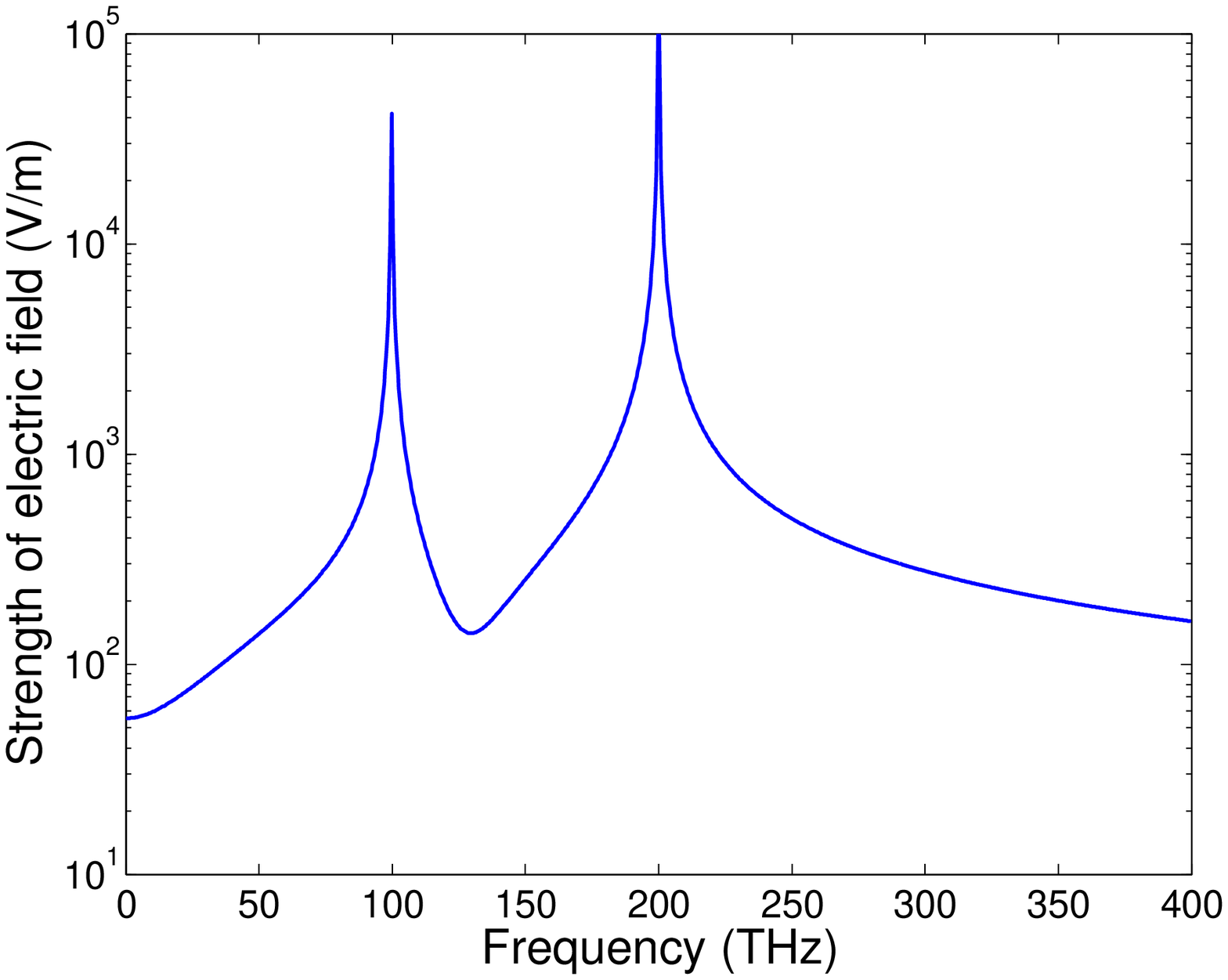}}
\caption {%
 (Color online)
 The transmitted waves and their corresponding Fourier transforms for different input powers. (a), (b) and (c) are the transmitted waves for input power $P_{\mathrm {in}} = 79.6,\ 90.7$ and  $120.6\,\mathrm {W/mm^2}$, respectively. (d), (e) and (f) are same as (a), (b) and (c), respectively, but for the Fourier transforms of the transmitted waves. The gain slab width $w=100\,\mathrm {nm}$ and the bandwidth $\Gamma_a$ of the atomic transition between $N_1$ and $N_2$ is $5\,\mathrm {THz}$.}
\label{fig3}
\end{figure*}

\begin{figure}
\center {
  \includegraphics[width=0.4\textwidth]{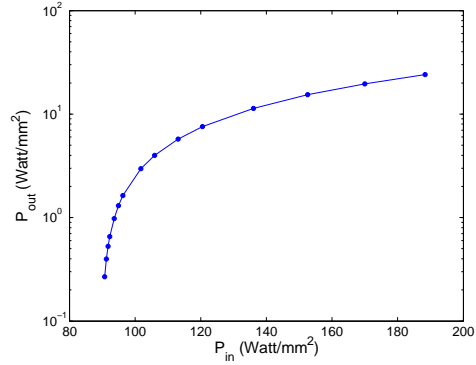}}
\caption {%
 (Color online)
 The powers emitted at the emission frequency $\omega = \omega_a$ ($100\,\mathrm {THz}$) for different input powers at the pumping frequency $\omega = \omega_b$ ($200\,\mathrm {THz}$). All parameters of this system are same as \fref {fig3}.}%
\label{fig4}
\end{figure}

To understand the lasing behavior of gain material, we first study a gain material slab surrounded by vacuum (shown in \fref {fig2}). We generate a continuous wave (CW) at the frequency $\omega_b$ ($200\,\mathrm {THz}$) and let it propagate through the gain slab, and then we calculate the reflected and transmitted waves and implement the Fourier transforms to see if there is lasing and how much power is emitted around $100\,\mathrm {THz}$ --- the emission frequency $\omega_a$ between $N_1$ and $N_2$. First we start with a very low input power $P_{\mathrm {in}}$ for the incident CW wave, but no lasing happens, then we increase the input power till it reaches the lasing threshold, for which the system starts to have lasing and we can see a small peak at the emission frequency $100 \, \mathrm {THz}$ in the Fourier transforms of reflected and transmitted waves, i.e., low power emitted around the emission frequency $\omega_a$. If we keep increasing the input power, the peak will get higher and the emitted power will get larger. \Fref {fig3} shows the transmitted waves and their corresponding Fourier transforms for three different input powers at the gain slab $w=100\,\mathrm {nm}$. We can see there is no lasing (\fref {fig3a}) when the input power is low ($P_{\mathrm {in}}=79.6\,\mathrm {W/mm^2}$) and there is only one peak for the pumping frequency in its Fourier transform(\fref {fig3d}). When the input power $P_{\mathrm {in}}=90.7\,\mathrm {W/mm^2}$, the system starts lasing (\fref {fig3b}) and a small peak appears at the frequency $100\,\mathrm {THz}$ (\fref {fig3e}). If we increase the input power to a higher value $P_{\mathrm {in}}= 120.6 \,\mathrm {W/mm^2}$, the lasing gets stronger (\fref {fig3c}) and the peak for the emission frequency gets higher (\fref {fig3f}), i.e., more power emitted around the emission frequency $\omega_a$. We have calculated the emitted power at the emission frequency $\omega_a$ versus the input power at the pumping frequency $\omega_b$ for the same gain slab system. As shown in \fref {fig4}, we can see that there is a sharp rise in the emission around $P_{\mathrm {in}} \approx 90.7\,\mathrm {W/mm^2}$, which corresponds to the lasing threshold for this system. Below the threshold, there is no lasing.

\begin{figure}
\center {
  \includegraphics[width=0.4\textwidth]{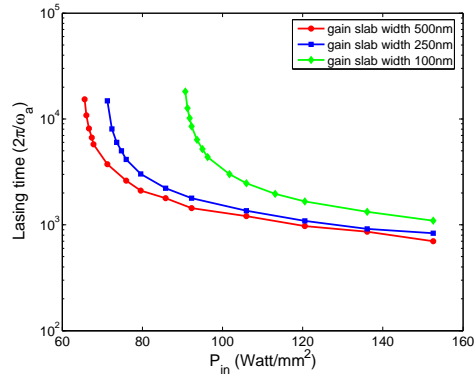}}
\caption {%
 (Color online)
 The lasing times for different input powers at the pumping frequency $\omega_b$ ($200\,\mathrm {THz}$). The gain slab width $w =100\,\mathrm {nm},\ 250\,\mathrm {nm}$ and $500\,\mathrm {nm}$, respectively. All other parameters are same as \fref {fig3}.}%
\label{fig5}
\end{figure}

We also notice that the lasing time (the time when the system starts lasing) varies according to the input power. \Fref {fig5} shows the detailed results for the lasing time versus the input power with the slab width $w=100\,\mathrm {nm},\ 250\,\mathrm {nm}$ and $500\,\mathrm {nm}$. We can see the lasing time decreases as the input power increases because the system pumps the electrons at a higher rate from the ground state level to higher levels and then reaches the population inversion between $N_1$ and $N_2$ in a shorter time. For a fixed input power, one can see the lasing time decreases as the gain slab width gets larger. This occurs because more input energy is absorbed and then converted to lasing by the wider gain slab. 

\begin{figure}
\center {
  \includegraphics[width=0.4\textwidth]{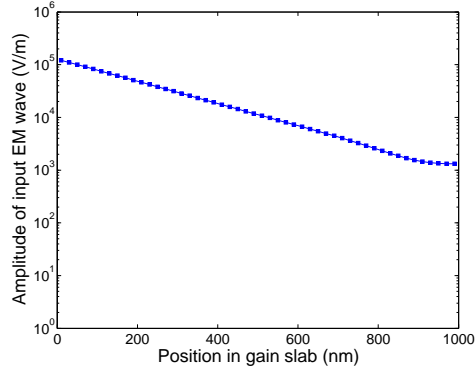}}
\caption {%
 (Color online)
 The amplitude of the input EM wave inside the gain slab as a function of the position. The gain slab width $w =1000\,\mathrm {nm}$ and the input power $P_{\mathrm {in}}=92.3\,\mathrm {W/mm^2}$. All other parameters are same as \fref {fig3}.}%
\label{fig6}
\end{figure}

\begin{figure}
\center {
  \includegraphics[width=0.4\textwidth]{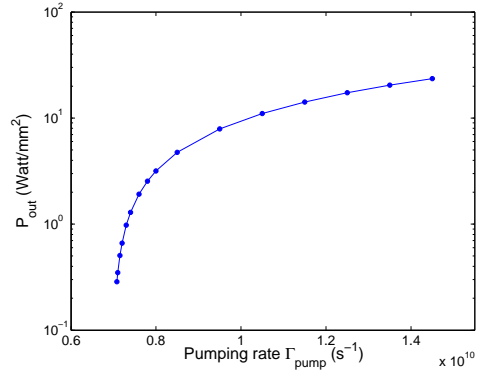}}
\caption {%
 (Color online)
The powers emitted at the emission frequency $\omega = \omega_a$ ($100\,\mathrm {THz}$) for different pumping rates. The gain slab width $w=100\,\mathrm {nm}$ and the bandwidth of the atomic transition between $N_1$ and $N_2$ is $5\,\mathrm {THz}$.}%
\label{fig7}
\end{figure}

\begin{figure*}
\centering
 \subfigure[]{
  \label {fig8a}
  \includegraphics[width=0.4\textwidth]{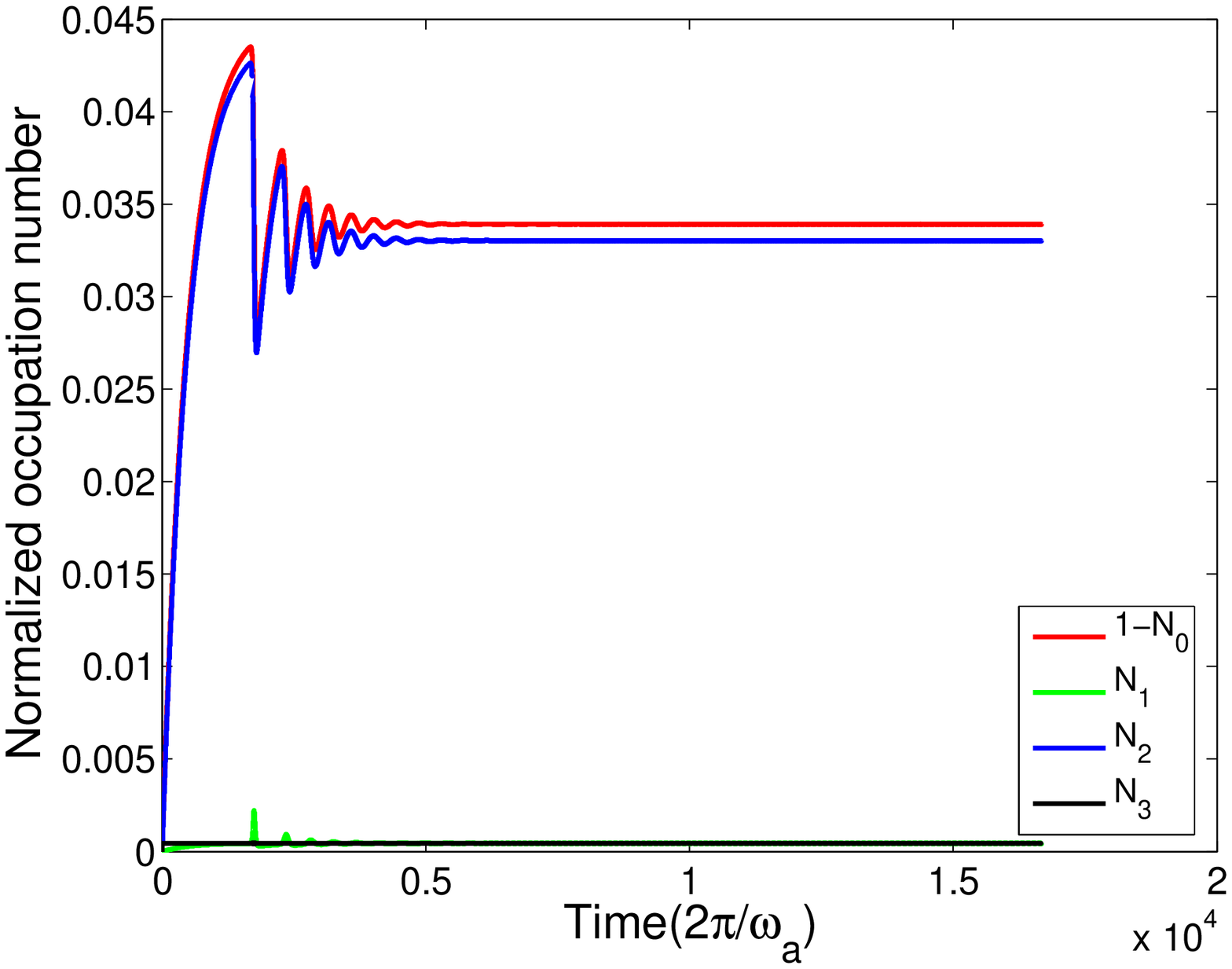}}
\qquad
\centering
 \subfigure[]{
  \label {fig8b}
  \includegraphics[width=0.4\textwidth]{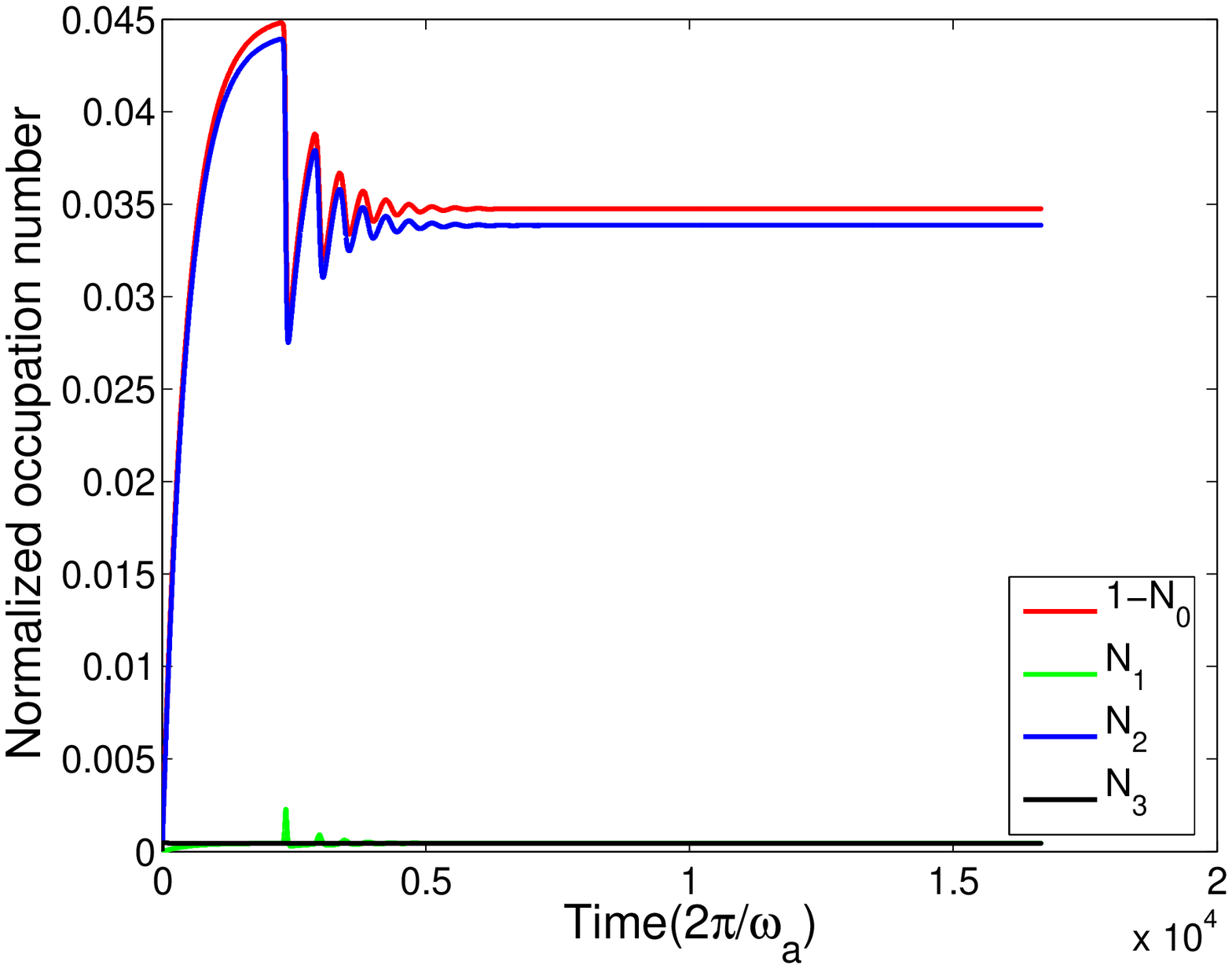}}
\caption {%
 (Color online) The normalized occupation numbers as a function of time. The gain slab width $w = 100\,\mathrm {nm}$ and the gain bandwidth of the atomic transition between $N_1$ and $N_2$ is $5\,\mathrm {THz}$. (a) the electrons are optically pumped by an input EM wave with input power $P_{\mathrm {in}}=120.6 \, \mathrm {W/mm^2}$ and (b) the electrons are pumped with a homogeneous pumping rate $\Gamma_{\mathrm {pump}}=9.3\times 10^9\,\mathrm {s^{-1}}$. Occupation numbers $N_0$, $N_1$, $N_2$ and $N_3$ are normalized by the total electron density $N_i$ [$N_i=N_0(t=0)=5.0\times 10^{23}\,\mathrm {/m^3}$].
}
\label {fig8}
\end{figure*}

As the input wave propagates inside the gain slab, it will decay due to the absorption from the gain material at the pumping frequency $\omega_b$ (see \fref {fig6}). Thus the pumping rate, which is determined by the local input optical intensity, is inhomogeneous inside the gain slab. But for a thin gain material layer, the electric field of the input wave can be approximately treated as homogeneous, thus we can simplify the pumping process between $N_0$ and $N_3$ by using a homogeneous pumping rate $\Gamma_{\mathrm {pump}}$. For a very thin gain slab $w=100\,\mathrm {nm}$, simulations are done with a homogeneous pumping rate and the results for the power emitted around the emission frequency $\omega_a$ versus the pumping rate are plotted in \fref {fig7}. Comparing with \fref {fig4}, where the electrons are optically pumped, we can see they are very similar. For a fixed output power such as $P_{\mathrm {out}}=7.56\,\mathrm {W/mm^2}$, we can find the corresponding input power $P_{\mathrm {in}}=120.6\,\mathrm {W/mm^2}$ in \fref {fig4} and the corresponding pumping rate $\Gamma_{\mathrm {pump}}=9.3\times 10^9\,\mathrm {s^{-1}}$ in \fref {fig7}. Then we do simulations for optical pumping case with the input power $P_{\mathrm {in}}=120.6\,\mathrm {W/mm^2}$ and for the homogeneous pumping rate case with the pumping rate $\Gamma_{\mathrm {pump}}=9.3\times 10^9\,\mathrm {s^{-1}}$. The graphs of the occupation numbers as a function time are plotted in \fref {fig8} for both cases. One can see that they are almost the same. This verifies that the homogeneous pumping rate simplification is valid for a thin gain slab. In our following simulations, we'll use this simplification because the gain slab widths in our structures are very thin ($w <=50\,\mathrm {nm}$). 

\subsection{Negative index material (NIM) embedded in layers of gain}\label {NIMInGain}

\begin{figure*}
\center {
  \includegraphics[width=0.6\textwidth]{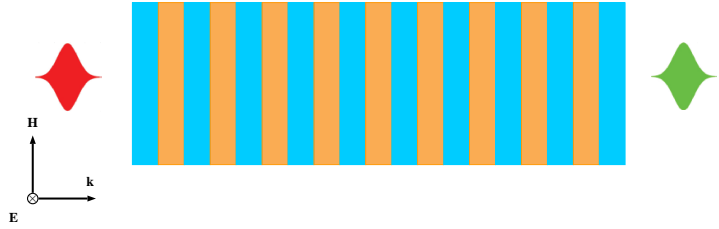}}
\caption {%
 (Color online)
 The negative index material (blue) embedded in layers of gain material (orange). The number of layers, the permittivity and permeability of NIM are taken different values for different cases we have examined. The width for both NIM and gain material is $w=50\,\mathrm {nm}$. The gain bandwidth is $5\,\mathrm {THz}$.}
\label{fig9}
\end{figure*}
\begin{figure}
\centering
  \includegraphics[width=0.4\textwidth]{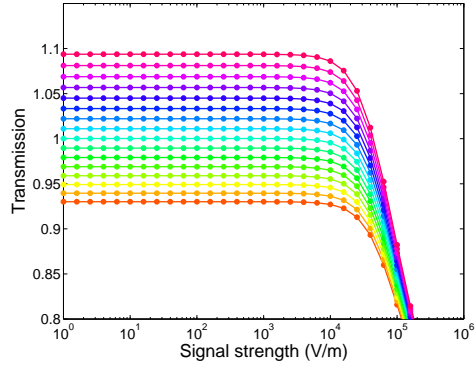}
\caption {%
 (Color online)
The transmission vs. signal amplitude for the loss-compensated metamaterial of a three-layer system (NIM - gain material - NIM) with gain bandwidth of $5\,\mathrm {THz}$, for different pumping rates $\Gamma_{\mathrm {pump}}$. $\Gamma_{\mathrm {pump}}$ is increased from $4.0\times 10^{9}\,\mathrm {s^{-1}}$ (lowest) to $5.5\times 10^{9}\,\mathrm {s^{-1}}$ (highest) in steps of $1.0\times 10^8\,\mathrm {s^{-1}}$. The material parameters for NIM are $\varepsilon = \mu = -1+2\,i$. The metamaterial response is linear in a very wide range. When the loss-compensated transmission reaches exactly unity, the pumping rate $\Gamma_{\mathrm {pump}}=4.70\times 10^9 \, \mathrm {s^{-1}}$, which is called the critical pumping rate. For incident fields stronger than $10^4\,\mathrm {V/m}$ the metamaterial behaves nonlinearly.}
\label{fig10}
\end{figure}
\begin{figure}
\centering
  \includegraphics[width=0.4\textwidth]{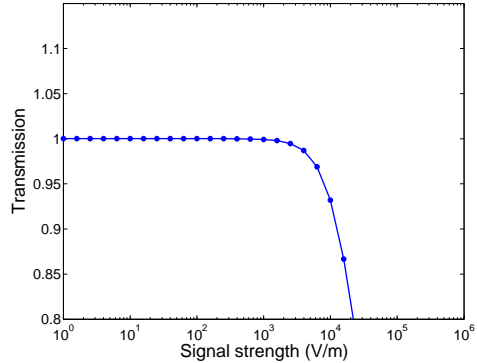}
\caption {%
 (Color online)
The transmission vs. signal amplitude for the loss-compensated metamaterial of \fref {fig9} with gain bandwidth of $5\,\mathrm {THz}$ at the critical pumping rates $\Gamma_{\mathrm {pump}}=1.98\times 10^9\,\mathrm {s^{-1}}$. The material parameters for NIM are same as \fref {fig10}. For incident fields stronger than $10^3\,\mathrm {V/m}$ this metamaterial becomes non-linear.}
\label {fig11}
\end{figure}

As the first simple model system, we will discuss a one-dimensional metamaterial system which consists of layers of negative index material (NIM) and gain material, as shown in \fref {fig9}, to see if we can compensate the losses of the metamaterials associated with the NIMs by the amplification provided by the gain material layers and how the system starts lasing. We first let a narrow band Gaussian pulse of a given amplitude go through the metamaterial without gain, and we calculate the transmitted signal emerging from the metamaterial system, which also has a Gaussian profile but its amplitude is much smaller than that of the incident pulse due to the losses of NIM layers. Then we introduce the gain into the system and start increasing the pumping rate. The amplitude of the transmitted signal gets larger and we can find a critical pumping rate, for which the transmitted pulse is of the same amplitude as the incident one. Since the gain material itself is nonlinear, we increase the amplitude of the incident Gaussian pulse for a fixed pumping rate to see how high we can go in the strength of the incident electric field and still have the full compensation of the losses, i.e., the transmitted signal equals the incident signal, independent on the signal strength. In this region we have compensated loss and still have linear response of the metamaterial. The shape of the transmitted signal is only affected by the dispersion but not dependent on the signal strength. For a three-layer system (NIM - gain material - NIM), we have calculated the transmission versus the strength of the electric field of the incident signal for several pumping rates close to the critical pumping rate $\Gamma_{\mathrm {pump}}= 4.70\times 10^9\,\mathrm {s^{-1}}$, as shown in \fref {fig10}. We found it has a linear response within a very broad range up to incident electric field of $10^3\,\mathrm {V/m}$. If we use 19 layers of \fref {fig9}, the critical pumping rate is $1.98\times 10^9\,\mathrm {s^{-1}}$, which is even smaller than half of the three-layer case, and the linear regime becomes narrower and drops faster than the 3-layer case for higher strength of incident electric field (shown in \fref {fig11}). To include the nonlinearity of gain material for strong incident signal, it is necessary to do a self-consistent calculation using FDTD method.

\begin{figure}
\centering
  \includegraphics[width=0.4\textwidth]{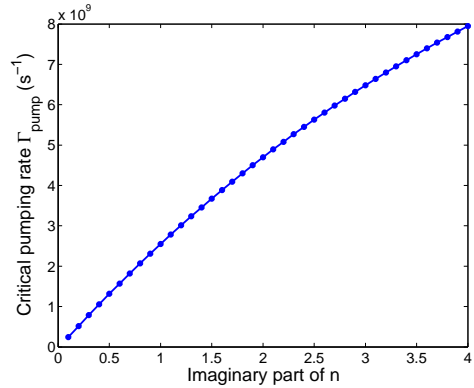}
\caption {%
 (Color online)
  The critical pumping rates for different imaginary parts of the refractive index $n$ of NIMs. The structure is a three-layer system (NIM - gain material - NIM) and $\varepsilon = \mu$ for NIMs.}
\label {fig12}
\end{figure}

As an example, we have also studied the three-layer system with different losses, to see how much $\Gamma_{\mathrm {pump}}$ we need to compensate the losses. \Fref {fig12} shows there exists a linear relation between the critical pumping rate and the imaginary part of the refractive index $n$ of NIMs.

\begin{figure*}
\centering
 \subfigure[]{
  \includegraphics[width=0.3\textwidth]{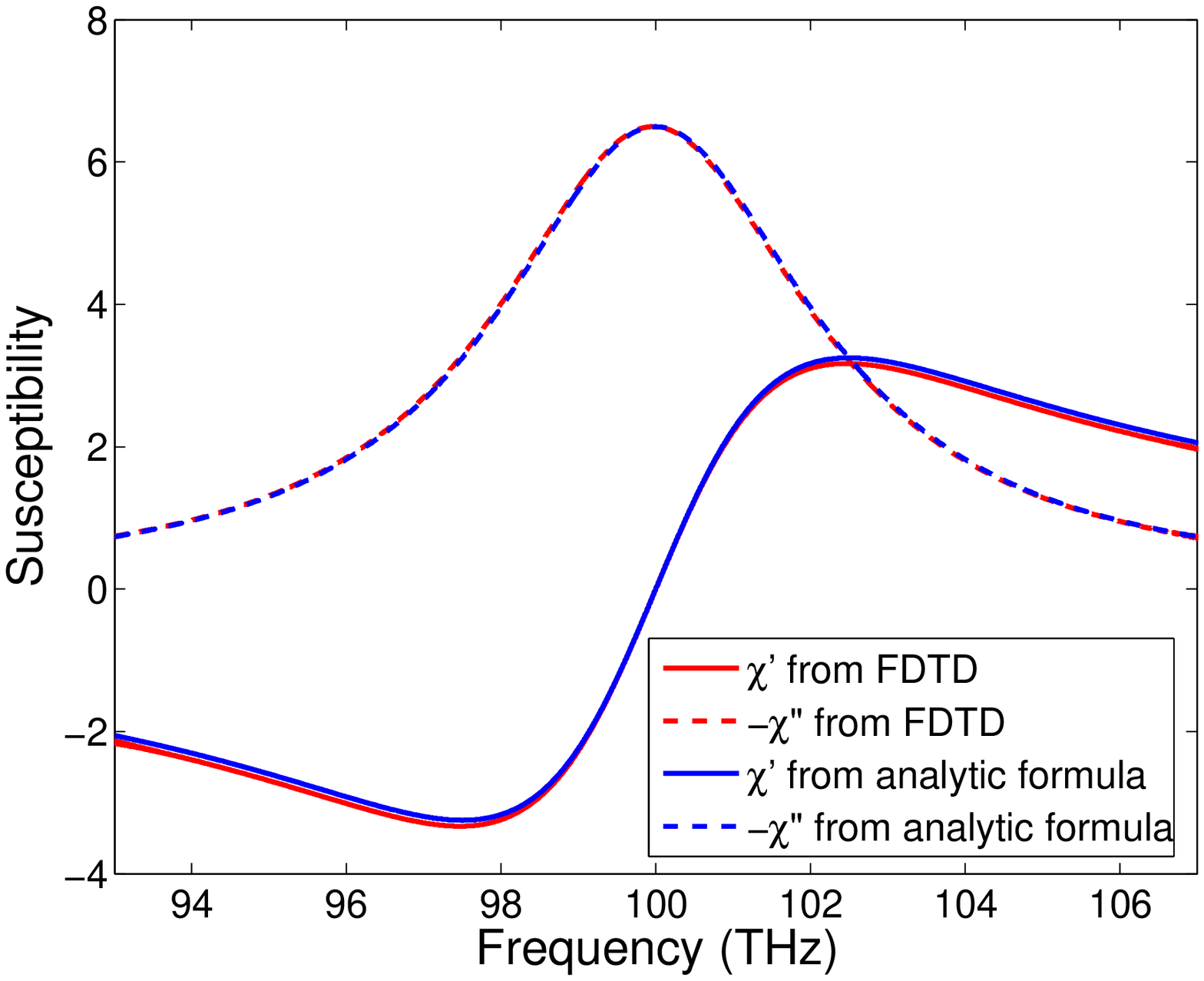}}
\qquad
\centering
 \subfigure[]{
  \includegraphics[width=0.3\textwidth]{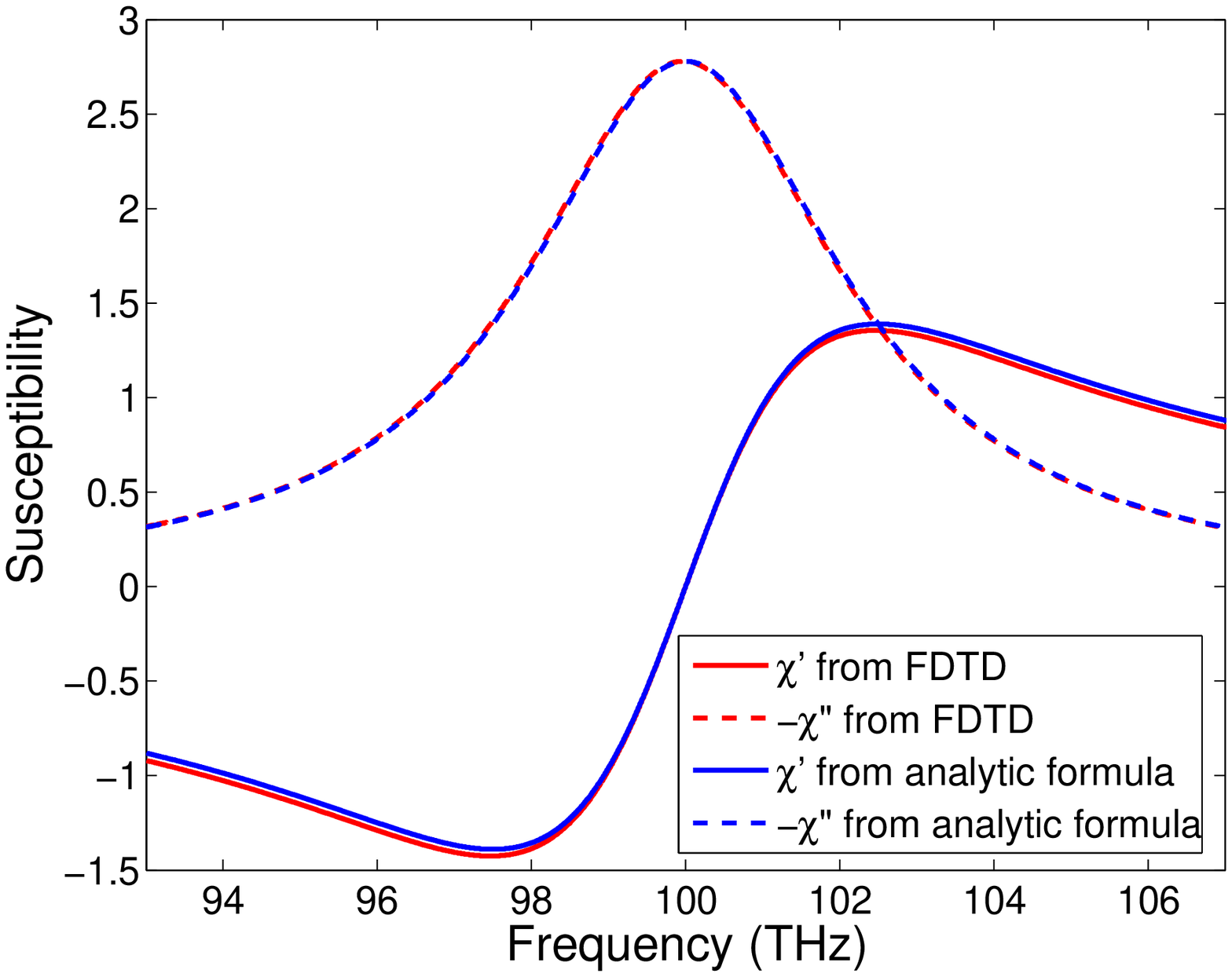}}
\caption {%
 (Color online)
 The numerical and analytical results for the susceptibilities of gain materials as a function of frequency. (a) a three-layer system (NIM - gain material -NIM) at the critical pumping rate $\Gamma_{\mathrm {pump}}=4.7\times 10^9\,\mathrm {s^{-1}}$. (b) a 19-layer system of \fref {fig9} at the critical pumping rate $\Gamma_{\mathrm {pump}}=1.98\times 10^9\,\mathrm {s^{-1}}$.}
\label {fig13}
\end{figure*}

We have also numerically calculated the susceptibilities of the gain material to see if it really has a Lorentzian lineshape. We first let a Gaussian pulse of a given amplitude ($10\,\mathrm {V/m}$) propagate through the metamaterial system and calculate the time-domain electric polarization $\mathbf {P}(\mathbf {r}, t)$ and the local electric field $\mathbf {E}(\mathbf {r},t)$. Then we implement the Fourier transforms to obtain the frequency-domain polarization and electric fields and calculate the frequency-dependent susceptibility by using the equation $\chi'(\omega)+i\chi''(\omega)=P(\omega)/\varepsilon_0 E(\omega)$. Simulations are done for both 3 and 19 layers and results are compared with the analytic results calculated using the equations \cite {14} $\chi' = -\chi_0'' \Delta x/(1+\Delta x^2)$ and $\chi'' = \chi_0'' /(1+\Delta x^2)$ with $\Delta x = 2(\omega-\omega_a)/\Gamma_a$ and $\chi_0''= -\sigma_a \Delta N/(\varepsilon_0^{}\omega_a\Gamma_a)$,  where $\Delta N = N_2-N_1$. As shown in \fref {fig13}, we found that the numerical susceptibilities are the same as the analytic ones and they do have a Lorentzian lineshape.

\begin{figure*}
\centering
 \subfigure[]{
 \label {fig14a}
  \includegraphics[width=0.3\textwidth]{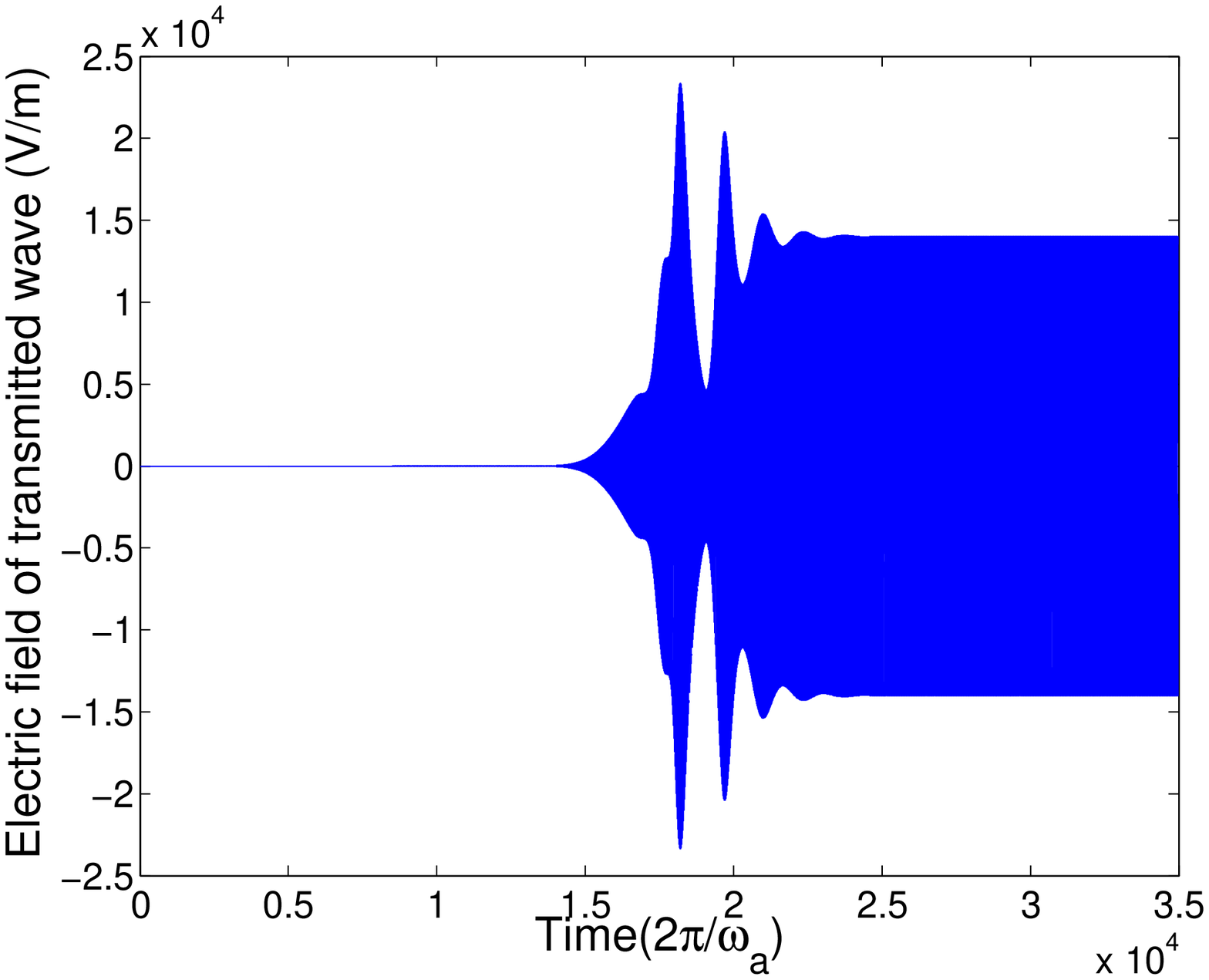}}
\qquad
\centering
 \subfigure[]{
  \label {fig14b}
  \includegraphics[width=0.3\textwidth]{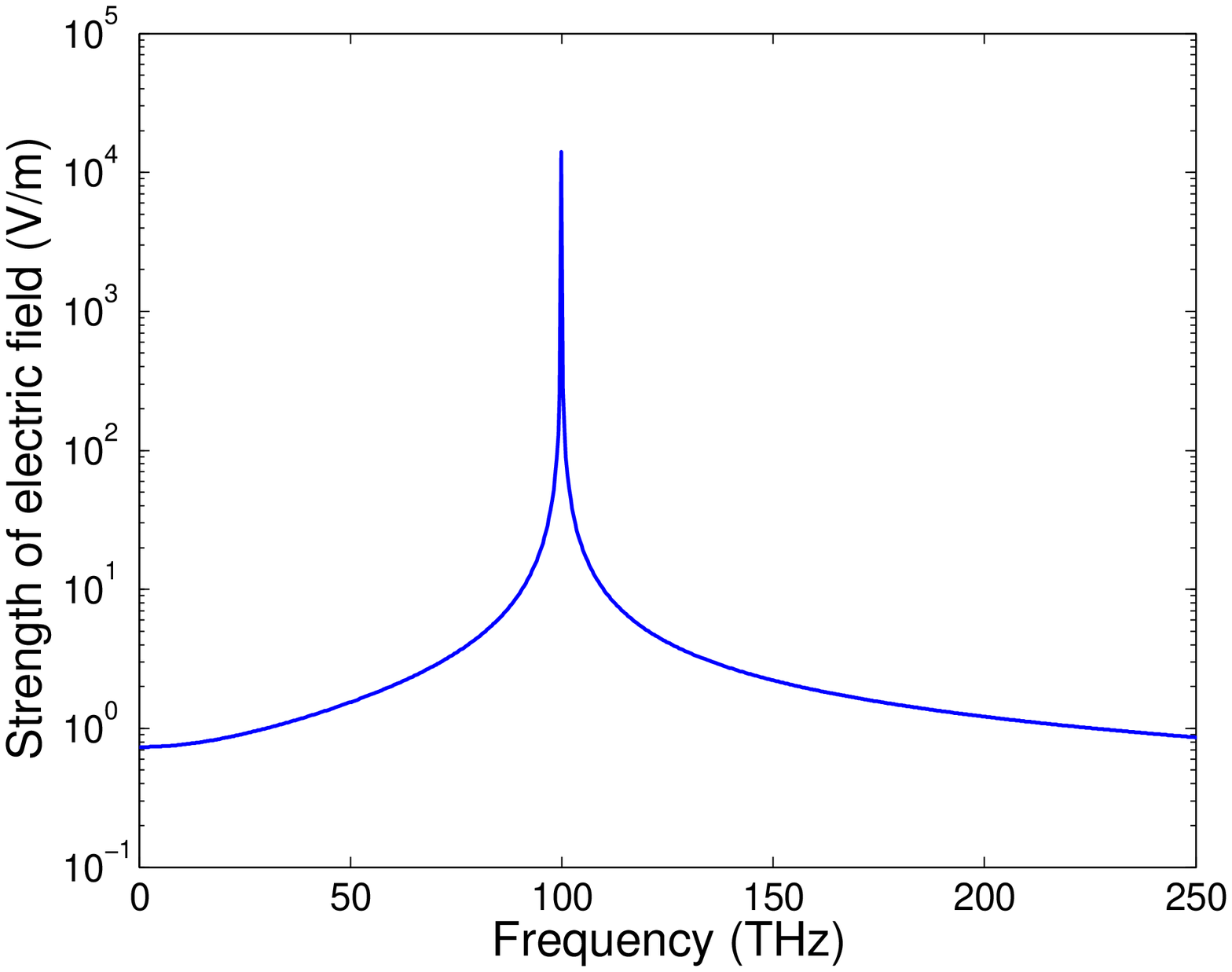}}
\caption {%
 (Color online)
 (a) the time-dependent electric field of the transmitted wave for a three-layer system (NIM - gain material - NIM) and (b) the corresponding Fourier transform in frequency domain for the lasing in (a). The pumping rate $\Gamma_{\mathrm {pump}}=1.5\times 10^{10} \,\mathrm {s^{-1}}$.}
\label {fig14}
\end{figure*}

To understand the lasing behavior of the metamaterial system, we increase the pumping rate to provide more gain from the gain material. We found the amplification of the incident signal gets larger and at last the system starts lasing till the pumping rate reaches a certain high value. \Fref {fig14a} shows the lasing behavior at the pumping rate $\Gamma_{\mathrm {pump}}=1.5\times 10^{10}\,\mathrm {s^{-1}}$ and there appears a peak at the emission frequency $100\,\mathrm {THz}$ in the corresponding Fourier transform (\fref {fig14b}).

If we use real metal layers instead of negative index materials, we can not compensate the losses of the metals. The reason is that the permittivity $\varepsilon$ for metals is large and negative and we'll have large reflections due to the impedance mismatch.

\subsection{One layer of gain material embedded in a square lattice of Lorentz dielectric cylinders}\label {CylinderInGain}

\begin{figure}
\centering
  \includegraphics[width=0.4\textwidth]{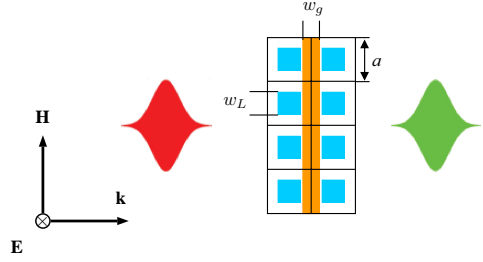}
\caption {%
 (Color online)
  One layer of gain material (orange) embedded in a square lattice of dielectric square cylinders (blue) that have a Lorentz behavior. The dielectric constant of the cylinders is given by $\varepsilon=1+\omega_p^2/(\omega_p^2-2i\omega\gamma-\omega^2)$, where the resonance frequency $f_p=\omega_p/2\pi=100\,\mathrm {THz}$ and $\gamma=2\pi f$, and $f$ takes different values in the cases we have examined. The dimensions are $a=80\,\mathrm {nm}$, $w_L=40\,\mathrm {nm}$, and $w_g = 30\,\mathrm {nm}$.}
\label {fig15}
\end{figure}

In \sref {NIMInGain}, we simply force the permittivity and the permeability of the metamaterial to be negative to have an unrealistic negative index material. In this section, we consider a two-dimensional (2D) metamaterial system (shown in \fref {fig15}) which consists of one layer of gain material and two layers of dielectric wires that have a Lorentz-type resonant electric response to emulate the resonant elements in a realistic metamaterial, such as cut-wires.

\begin{figure*}
\centering
 \subfigure[]{
 \label {fig16a}
  \includegraphics[width=0.3\textwidth]{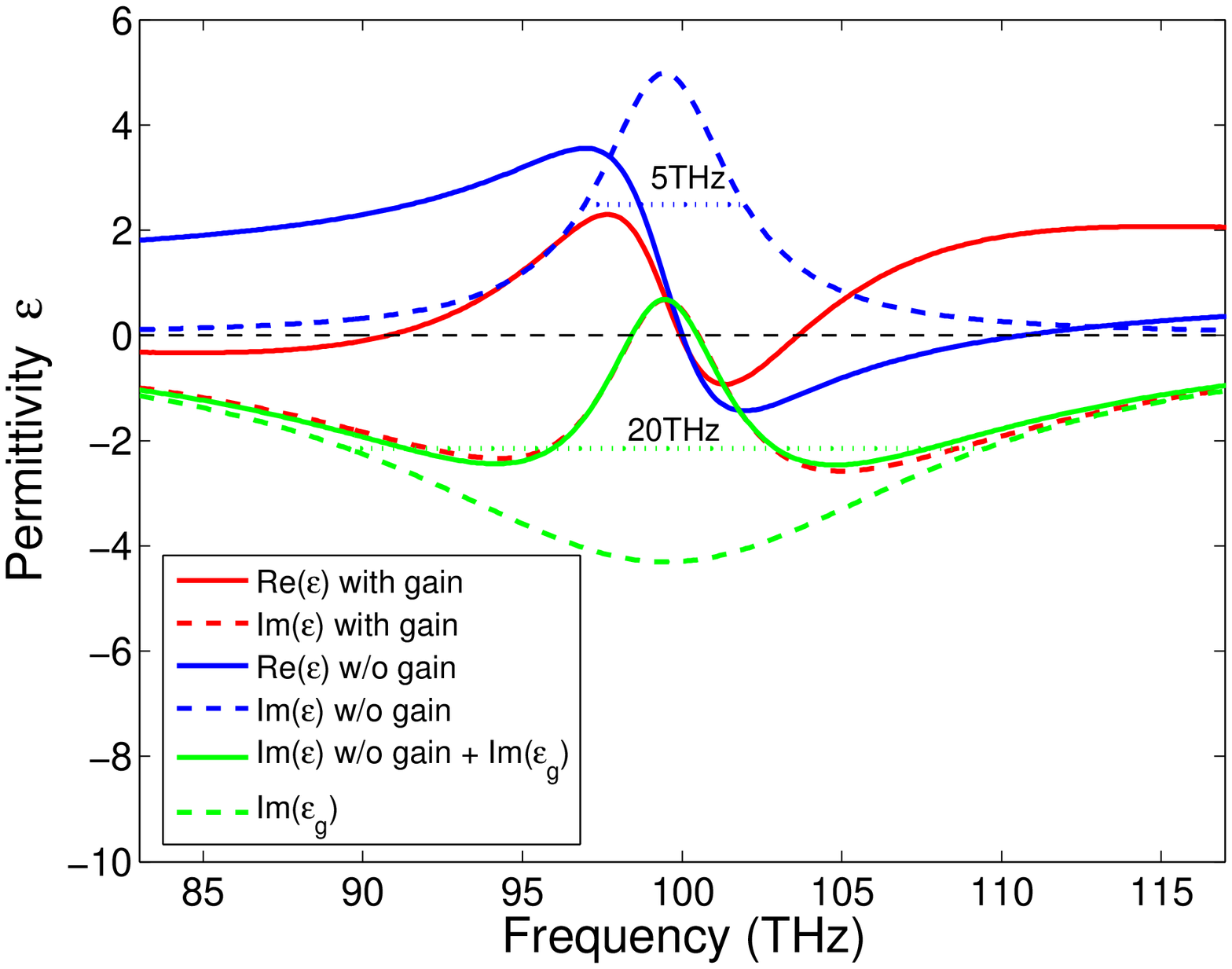}}
\centering
 \subfigure[]{
 \label {fig16b}
  \includegraphics[width=0.3\textwidth]{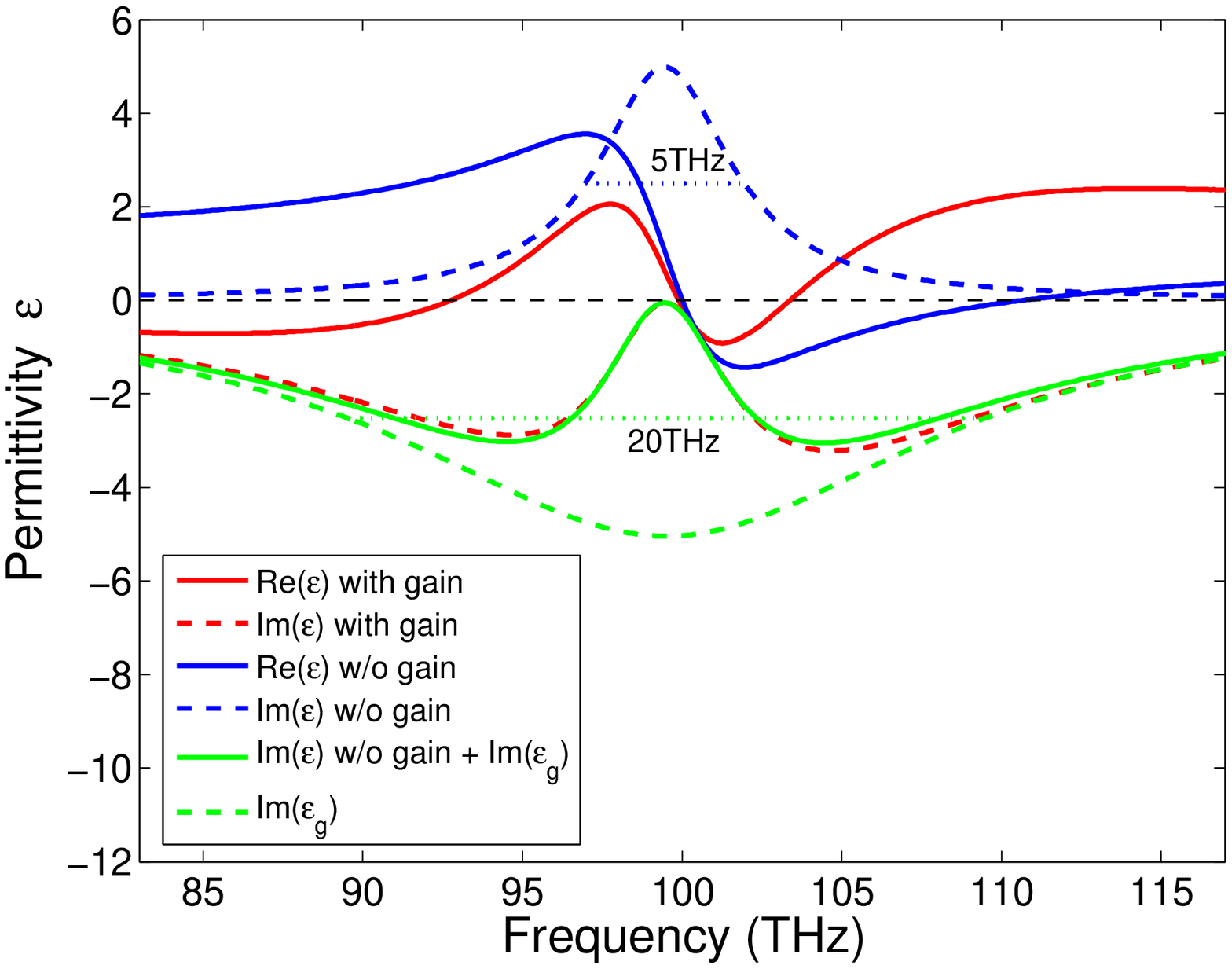}}
\centering
 \subfigure[]{
 \label {fig16c}
  \includegraphics[width=0.3\textwidth]{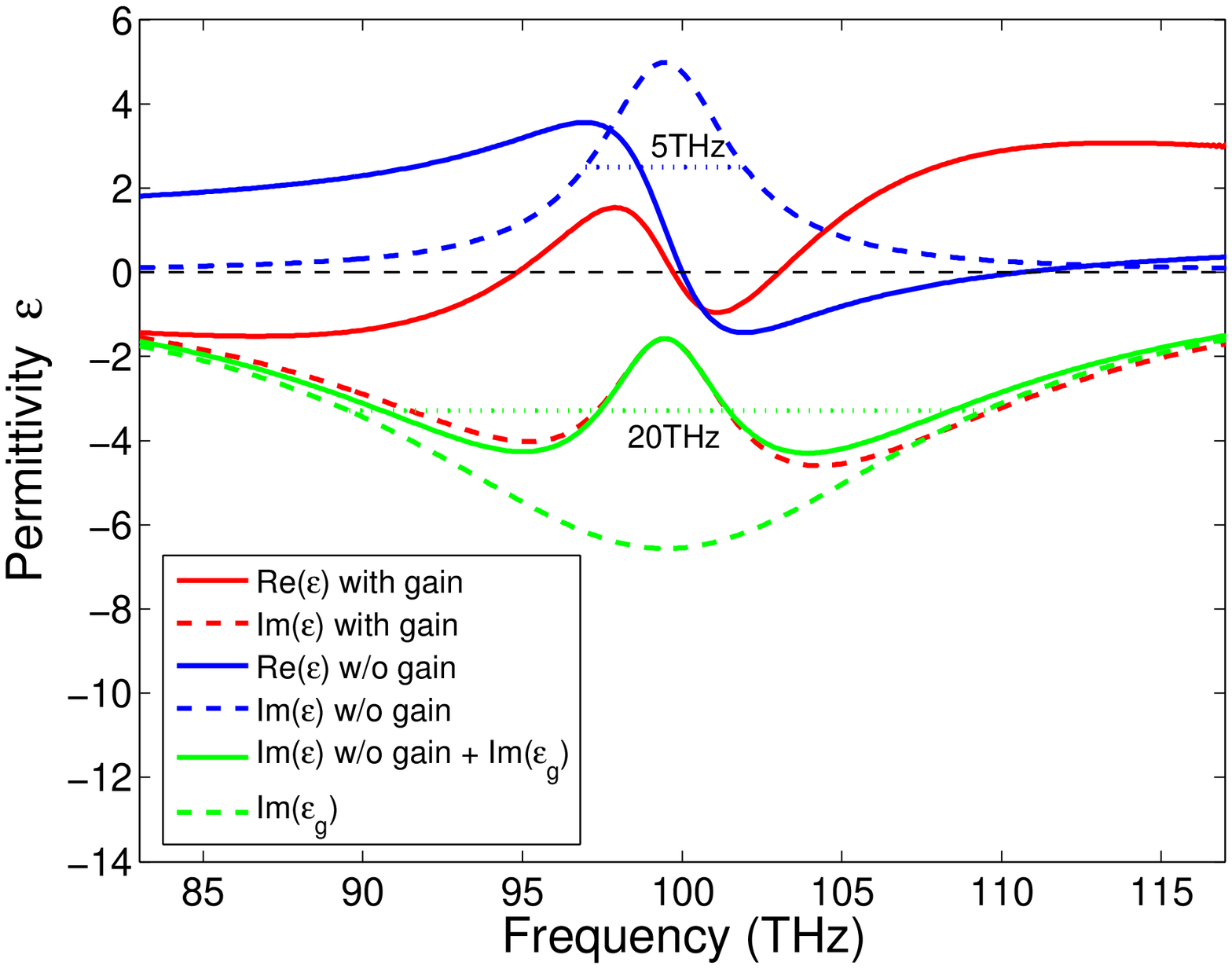}}
\\
\centering
 \subfigure[]{
  \label {fig16d}
  \includegraphics[width=0.3\textwidth]{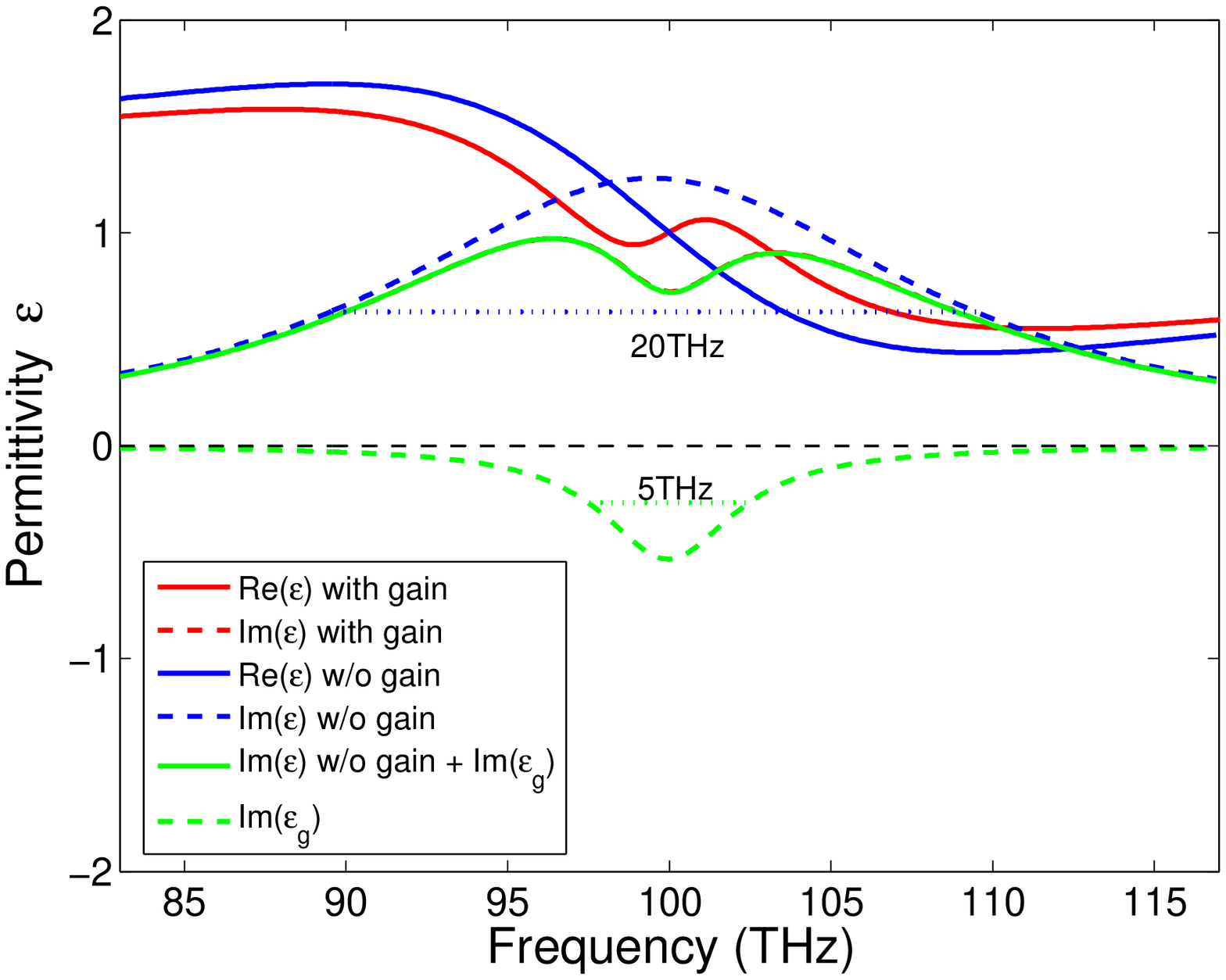}}
\centering
 \subfigure[]{
 \label {fig16e}
  \includegraphics[width=0.3\textwidth]{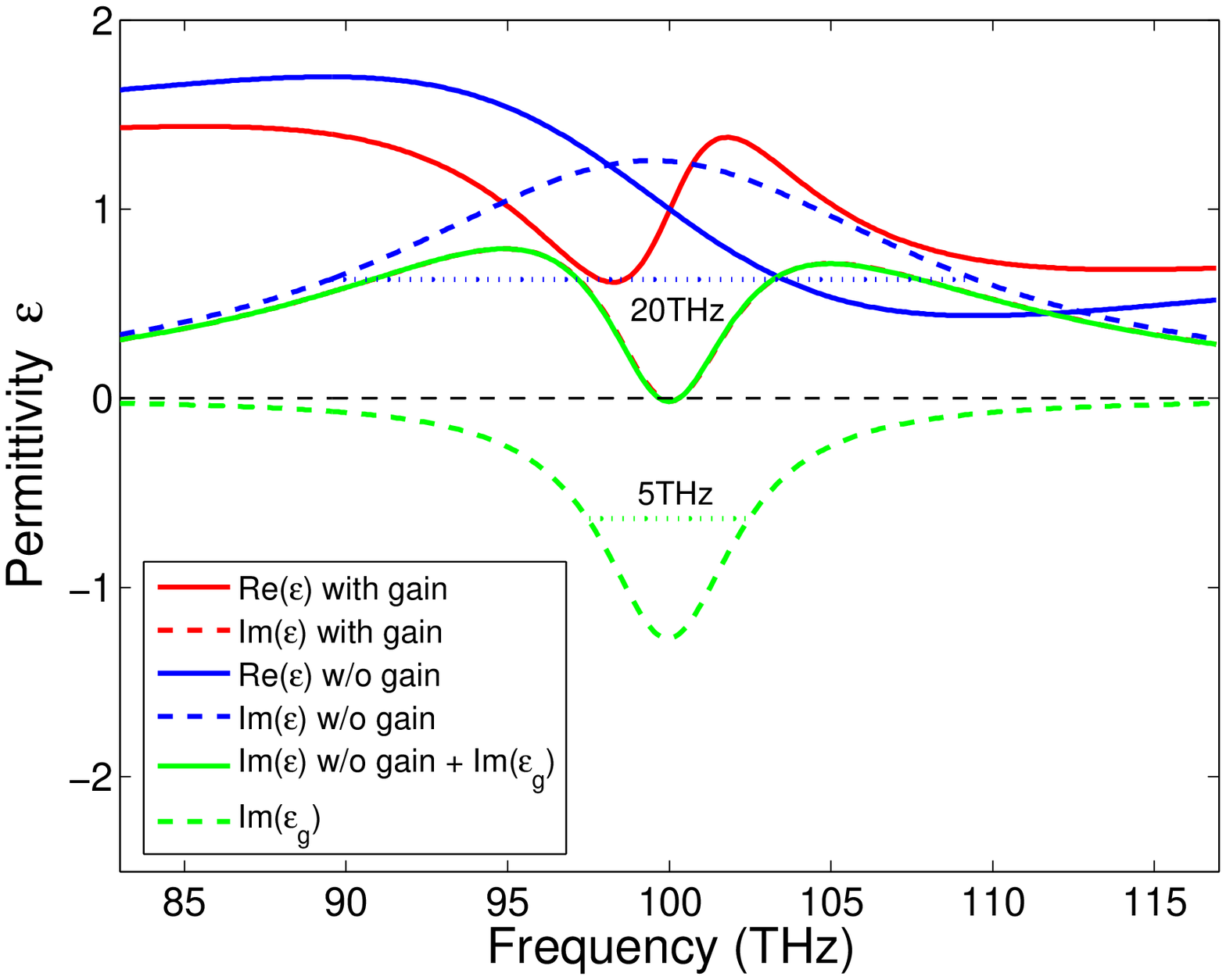}}
\centering
 \subfigure[]{
 \label {fig16f}
  \includegraphics[width=0.3\textwidth]{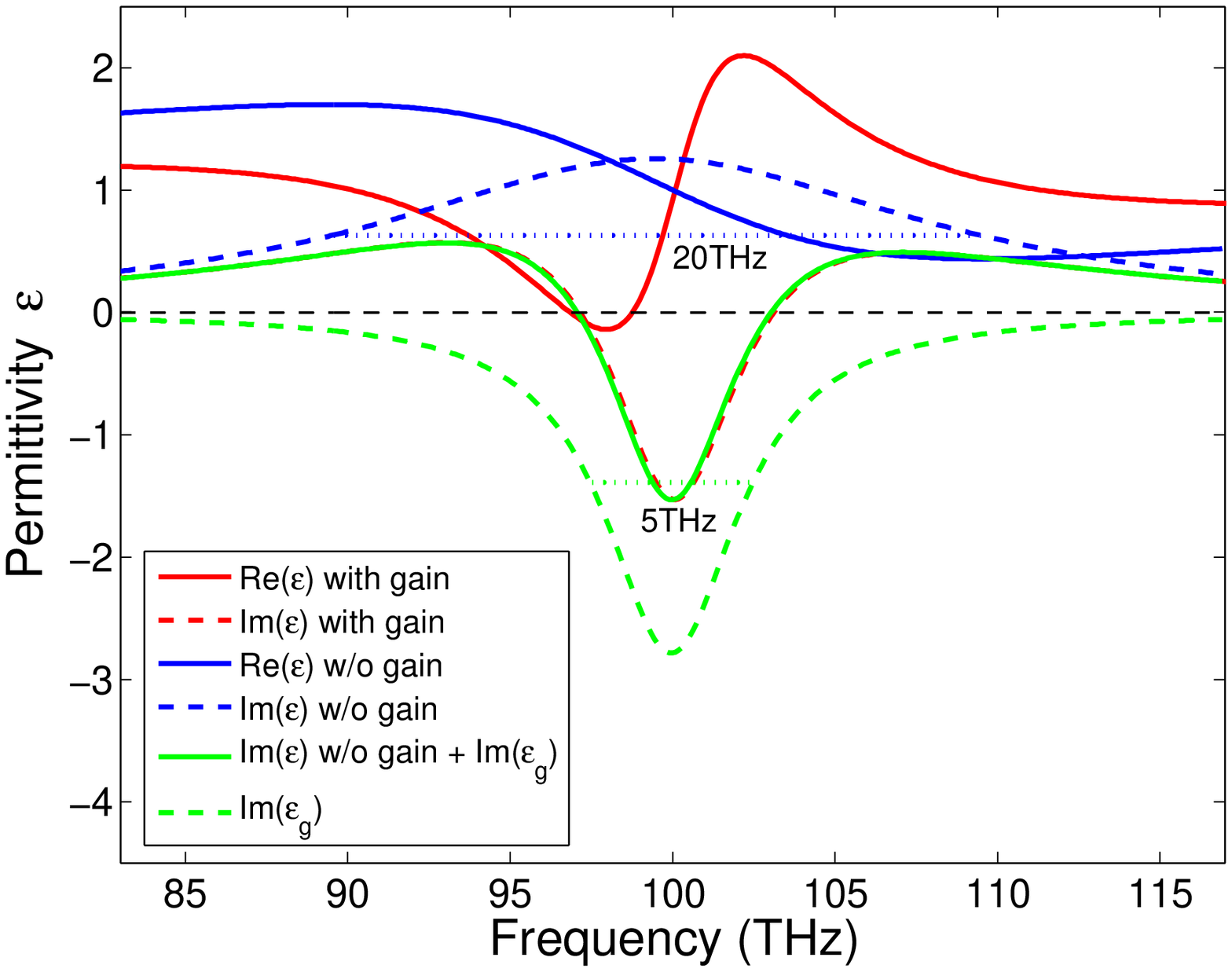}}
\caption {%
 (Color online)
 (a) The retrieved results for the real and the imaginary parts of the effective permittivity $\varepsilon$ with gain and without gain. Below compensation, $t=0.89$; gain and Lorentz bandwidths are $20\,\mathrm {THz}$ and $5\,\mathrm {THz}$, respectively. (b) and (c) are same as (a), but for the loss compensated case ($t=1$) and the overcompensated case ($t=1.34$), respectively. (d) - (f) are same as (a) - (c), respectively, except with gain bandwidth $5\,\mathrm {THz}$ and Lorentz bandwidth $20\,\mathrm {THz}$.}
\label {fig16}
\end{figure*}

We first study the three-layer system of \fref {fig15} with the full width at half maximum (FWHM) $5\,\mathrm {THz}$ for Lorentz dielectric (i.e., $f=2.5\,\mathrm {THz}$) and $20\,\mathrm {THz}$ for gain. The transmission, $T$, the reflection, $R$, and the absorption, $A=1-T-R$, as a function of frequency for the system are obtained in the propagation direction. With the introduction of gain, the absorption at the resonance frequency of $100\,\mathrm {THz}$ decreases and reaches 0 at a certain pumping rate. So the gain compensates the losses. If we continue increasing the gain, i.e., the pumping rate, the system gets overcompensated and the absorption becomes negative. To see how the losses of the emulated resonators get compensated by the gain, we exploit the usual retrieval procedure based on inverting the scattering amplitudes \cite {18} to obtain the effective permittivities $\varepsilon$ without gain and with gain. \Fref {fig16}(a) - 16(c) show the retrieved results for the real and the imaginary parts of the effective permittivities $\varepsilon$ of the system for the below compensation, loss-compensated, and overcompensated cases, respectively, together with the effective permittivity without gain. The retrieved results for $\varepsilon$ without gain have exactly the Lorentzian shape but the amplitude of the real and the imaginary parts of $\varepsilon$ is a factor of 4 less than the Lorentz formula for the square cylinders. This is due to the filling ratio of the square cylinders in the unit cell. Due to the loss compensation from the gain material, one can see the imaginary part of the effective permittivity gets lower as the gain increases. Below compensation, its value at the resonance frequency of $100\,\mathrm {THz}$ is  positive, while it's zero and negative for the loss-compensated and overcompensated cases, respectively. Notice that we can have $\mathrm {Re}(\varepsilon)\approx 2$ with $\mathrm {Im}(\varepsilon) \approx 0$ at $98\,\mathrm {THz}$ and $\mathrm {Re}(\varepsilon)\approx -1$ with $\mathrm {Im}(\varepsilon) \approx 0$ at $101\,\mathrm {THz}$ for the below compensation case (\fref {fig16a}). For the loss-compensated case (\fref {fig16b}), we have $\mathrm {Re}(\varepsilon)\approx 1$ with $\mathrm {Im}(\varepsilon) \approx 0$ at the resonance frequency $100\,\mathrm {THz}$, which is just same as vacuum and makes no sense for us. For the overcompensated case (\fref {fig16c}), the imaginary part of the effective permittivity $\varepsilon$ is negative within all the frequency range.

Second, we study the three-layer system with the FWHM of gain smaller than Lorentz dielectric, where the bandwidths for gain and Lorentz dielectric are $5\,\mathrm {THz}$ and $20\,\mathrm {THz}$ (i.e., $f=10\,\mathrm {THz}$), respectively. The introduction of gain develops a peak at the resonance frequency of $100\,\mathrm {THz}$ for the transmission while the absorption has a dip. The retrieved results for the real and the imaginary parts of the effective permittivities $\varepsilon$ without gain and with gain are plotted in \fref {fig16d} - 16(f) for the three different cases discussed above. Similar to the first case we examined where the loss bandwidth is smaller than the gain, the imaginary part of the effective permittivity $\varepsilon$ gets smaller due to the gain. The difference is we get interesting results for the overcompensated case instead of the below compensation case, where we can have $\mathrm {Re}(\varepsilon)\approx 0$ with $\mathrm {Im}(\varepsilon)\approx 0$ at $97\,\mathrm {THz}$ and $\mathrm {Re}(\varepsilon)\approx 2.1$ with $\mathrm {Im}(\varepsilon)\approx 0$ at $103\,\mathrm {THz}$.

So for both the two systems, one can obtain a lossless metamaterial with positive or negative $\mathrm {Re}(\varepsilon)$, either below compensation or over compensation. In \fref {fig16} we also have plotted the sum of $\mathrm {Im}(\varepsilon)$ without gain and the imaginary part of $\varepsilon_g$, the dielectric function of the gain material. One can see the imaginary part of $\varepsilon$ of our total system with gain is equal to the sum of $\mathrm {Im}(\varepsilon)$ and  $\mathrm {Im}(\varepsilon_g)$. This is unexpected because there is no coupling between the Lorentz dielectric and the gain. This is indeed true for the 2D Lorentz dielectric cylinders, because they have a continuous shape like a solenoid and the gain material slabs have zero depolarization field. Different from finite length wires [hence a three-dimensional (3D) problem] where the dipole interactions between Lorentz wires and gain material are dominated by the near field $O(1/r^3)$, the interaction for infinite length wires is only via the propagating field $O(\omega\ln|kr|)$, and much weaker. That's why the Lorentz wires and the gain material are approximately independent in our 2D simulations. So there is a need for a true 3D simulation to solve this problem and obtain different behaviors. However, the 3D simulation is computationally excessively demanding.

Like the layered system in \sref {NIMInGain}, if we keep increasing the pumping rate, i.e., the gain, at last both of the two systems will have lasing. For example, when the pumping rate reaches $\Gamma_{\mathrm {pump}}=3.2\times 10^{10}\,\mathrm {s^{-1}}$, the three-layer system with gain bandwidth $5\,\mathrm {THz}$ and Lorentz bandwidth $20\,\mathrm {THz}$ starts lasing --- the system itself has a coherent self-sustained steady output. 
 
\subsection{2D split ring resonators (SRRs) with gain material inclusions}\label {SRRInGain}

\begin{figure}
\centering
  \includegraphics[angle=-90, width=0.4\textwidth]{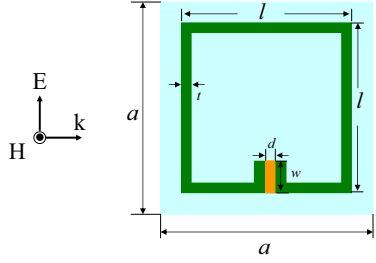}
\caption {%
 (Color online)
 Geometry for a unit cell of the square SRR system with gain embedded in the gap (orange). The dimensions are $a=100\,\mathrm {nm}$, $l=80\,\mathrm {nm}$, $t=5\,\mathrm {nm}$, $d=5\,\mathrm {nm}$ and $w=15\,\mathrm {nm}$.}
\label {fig17}
\end{figure}

To avoid the decoupling problem in \sref {CylinderInGain} and still limit our simulations in a 2D model, we consider a 2D split ring resonator (SRR) as a more realistic and also more relevant model, where the relevant polarization is across the finite SRR gap and therefore the coupling to the gain material is dipolelike. In \fref {fig17}, we show the geometry for the unit cell of the SRR system with the gain material embedded in the gap. The dimensions of the SRR are chosen such that the magnetic resonance frequency of the SRR overlaps with the emission frequency ($100\,\mathrm {THz}$) of the gain material. Due to the strong electric field inside the gap, there will be strong coupling between the SRR and the gain material. We also want to see if the losses of the magnetic response can be compensated by the electric gain. 

\begin{figure*}
\centering
 \subfigure[]{
 \label {fig18a}
  \includegraphics[width=0.3\textwidth]{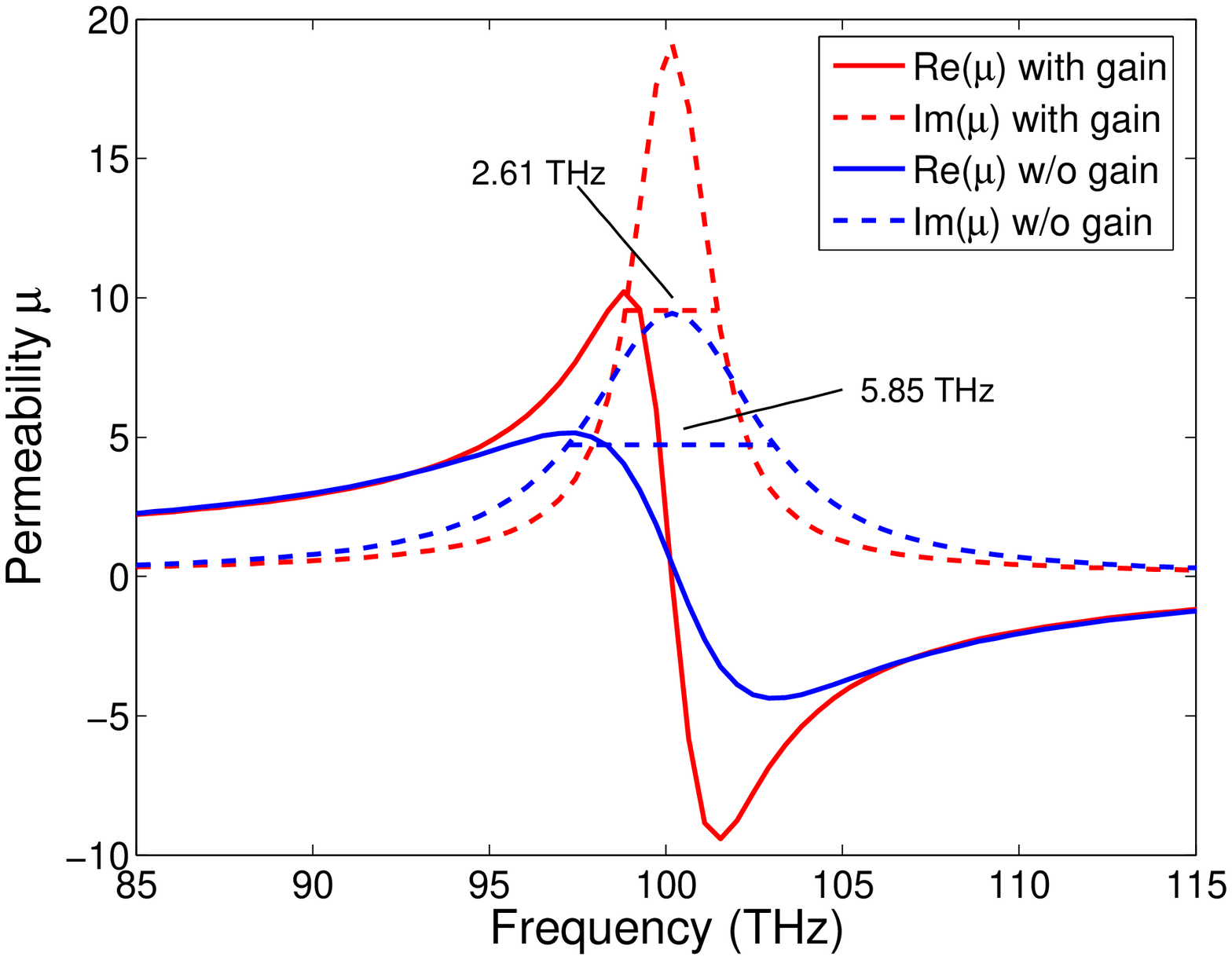}}
\qquad
\centering
 \subfigure[]{
  \label {fig18b}
  \includegraphics[width=0.3\textwidth]{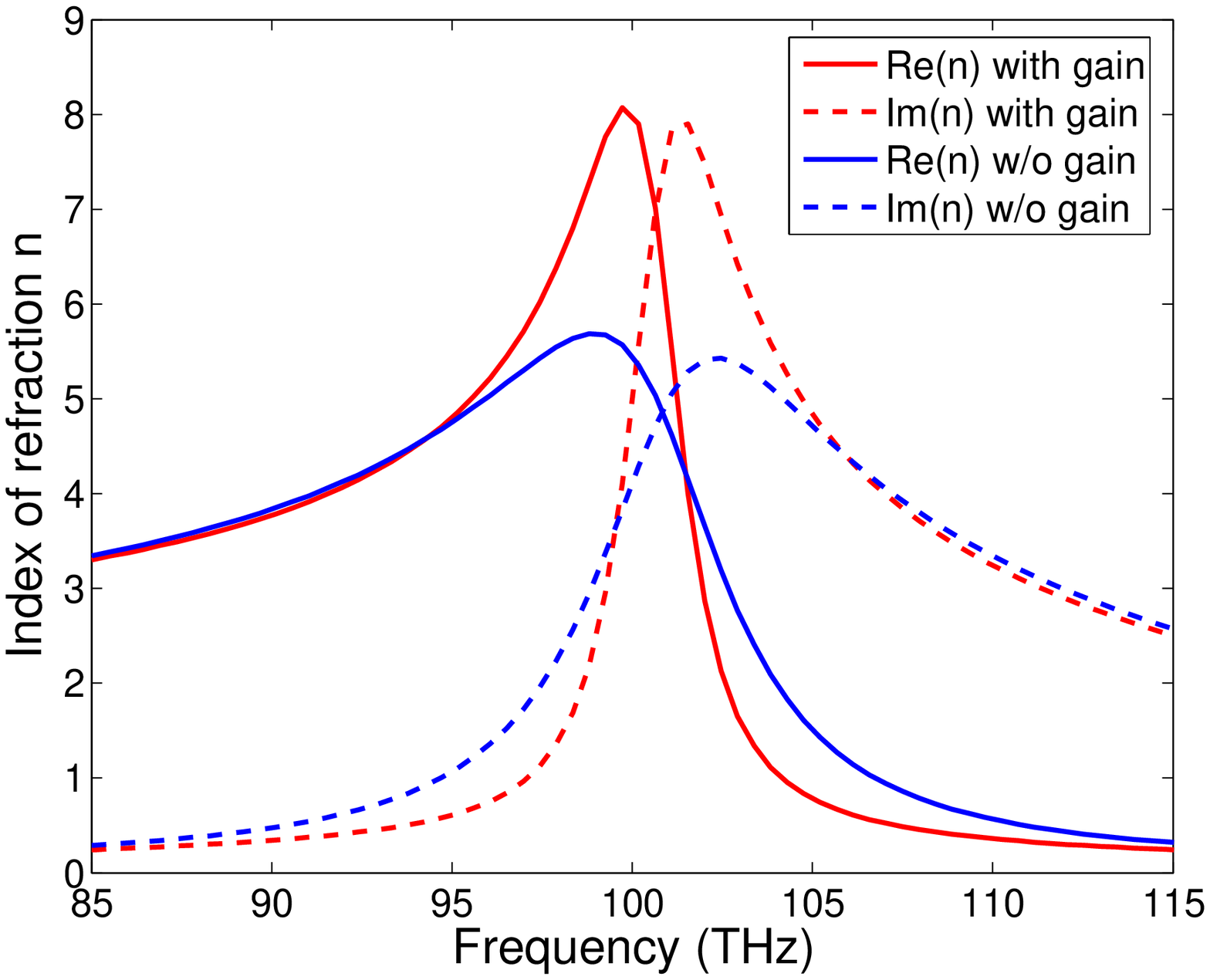}}
\caption {%
 (Color online)
 The retrieved results for the real and the imaginary parts of (a) the effective permeability $\mu$ and (b) the corresponding effective index of refraction $n$, with and without gain for a pumping rate $\Gamma_{\mathrm {pump}}=1.0\times 10^9\,\mathrm {s^{-1}}$ and the SRR system of \fref {fig17}. The gain bandwidth is $20\,\mathrm {THz}$. Notice that the width of the magnetic resonance with gain is $2.61\,\mathrm {THz}$.}
\label {fig18}
\end{figure*}

Simulations are done for one layer of the square SRR with gain bandwidth of $20\,\mathrm {THz}$. \Fref {fig18a} shows the retrieved results for the real and the imaginary parts of the effective permeability $\mu$, with gain (pumping rate $\Gamma_{\mathrm {pump}}=1.0\times 10^9\,\mathrm {s^{-1}}$) and without gain. One can see that with the introduction of gain, the gain undamps the magnetic resonance of the SRR and the weak and broad resonant effective permeability $\mu$ of the lossy SRR becomes strong and narrow. The FWHM with gain is $2.61\,\mathrm {THz}$, while the FWHM without gain is $5.85\,\mathrm {THz}$, which is more than twice larger than the former. Notice that in the off-resonance range in \fref {fig18a}, we can obtain the effective permeability $\mu$ with a smaller imaginary part with the introduction of the gain, which means the magnetic loss is compensated by the electric gain. \Fref {fig18b} shows the retrieved results for the real and the imaginary parts of the corresponding effective index of refraction $n$, with and without gain. Note that for a lossless SRR, $n$ is purely real away from the resonance except in a small band above the resonance where it's purely imaginary due to the negative $\mu$. At the frequency of $96\,\mathrm {THz}$, slightly below the resonance (\fref {fig18b}), the imaginary parts of the index of refraction $n$ without and with gain are $1.36$ and $0.754$. Then  we can find the effective extinction coefficient without gain is $\alpha =(\omega/c)\mathrm {Im}(n) \approx 2.74\times 10^4 \,\mathrm {cm^{-1}}$ and the one with gain is $\alpha \approx 1.52\times 10^4\,\mathrm {cm^{-1}}$. And hence the effective amplification coefficient of the gain in the combined system is $\alpha \approx -1.22 \times 10^4\,\mathrm {cm^{-1}}$, which is much larger than the amplification $\alpha \approx -9.2\times 10^2\,\mathrm {cm^{-1}}$  for the bulk gain material \cite {14} at the given pumping rate $\Gamma_{\mathrm {pump}}=1.0\times 10^9\,\mathrm {s^{-1}}$. This is due to the strong local electric field enhancement in the gap of the resonant SRR. While we have the incident electric field $10\,\mathrm {V/m}$, the induced electric field in the gap is around $450\,\mathrm {V/m}$. Taking the observed field enhancement factor around 45 in the gap of SRR, the energy produced by the gain in the gap is around $12$ times larger than by the homogeneous bulk gain material in the size of the unit cell, which agrees well with the factor around 15 between the simulated SRR effective medium and the homogeneous gain medium. If we continue increasing the pumping rate, the magnetic resonance becomes narrower ($0.96\,\mathrm {THz}$ for pumping rate $\Gamma_{\mathrm {pump}}=1.8\times 10^9\,\mathrm {s^{-1}}$). When the pumping rate reaches $\Gamma_{\mathrm {pump}}=1.9\times 10^9\,\mathrm {s^{-1}}$, the metamaterial system gets overcompensated and the imaginary part of the effective permeability $\mu$ at the resonance frequency gets flipped down and becomes negative. If we increase the pumping rate even more (around $\Gamma_{\mathrm {pump}}=5.0\times 10^9\,\mathrm {s^{-1}}$), the SRR system starts lasing \cite {19,20}.

\begin{figure}
\centering
  \includegraphics[ width=0.4\textwidth]{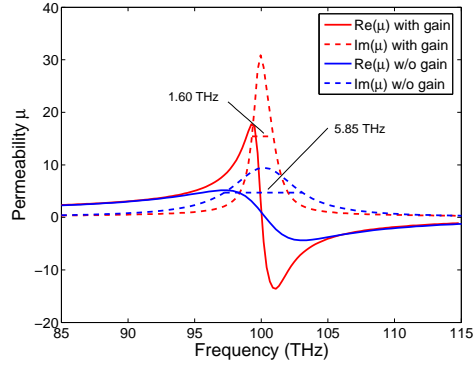}
\caption {%
 (Color online)
 The retrieved results for the real and the imaginary parts of the effective permeability $\mu$ with and without gain for a pumping rate $\Gamma_{\mathrm {pump}}=8.0\times 10^8\,\mathrm {s^{-1}}$ and the SRR system where the SRR is surrounded by gain. The gain bandwidth is $20\,\mathrm {THz}$. Notice that the width of the magnetic resonance with gain is $1.60\,\mathrm {THz}$, narrower than \fref {fig18a}.} 
\label {fig19}
\end{figure}
Instead of having the gain in the gap of SRR, we also have done simulations on the SRR system where the gain is surrounding around the SRR. The retrieved results for the real and the imaginary parts of the effective permeability $\mu$ without gain and with gain for a given pumping rate $\Gamma_{\mathrm {pump}}=8.0\times 10^8\,\mathrm {s^{-1}}$ are plotted in \fref {fig19}. Due to the larger gain filling in the unit cell, the coupling between the gain material and the SRR gets stronger and the losses of SRR are easier to be compensated. Compared with the case with the gain in the gap only, the effective permeability $\mu$ of the simulated SRR system in \fref {fig19} can have stronger and narrower resonance (FWHM = $1.6\,\mathrm {THz}$) even with lower pumping rate. Similar to the previous case, the magnetic resonance becomes stronger and narrower ($0.7\,\mathrm {THz}$ at $\Gamma_{\mathrm {pump}}=1.0\times 10^9\,\mathrm {s^{-1}}$) as we increase the pumping rate, and the system gets overcompensated so that the magnetic resonance peak gets flipped down when the pumping rate reaches $\Gamma_{\mathrm {pump}}=1.1\times 10^9\,\mathrm {s^{-1}}$. It's also easier for this system to have lasing \cite {19,20}, which is observed when the pumping rate is $\approx \Gamma_{\mathrm {pump}}=2.5\times 10^9\,\mathrm {s^{-1}}$.

\section{Conclusions} \label {conclusion}
We have performed numerical simulations on one dimensional gain material system by using FDTD method. The system starts lasing when the input reaches over the lasing threshold. For more input power and wider gain slab, the lasing is faster. Comparisons were done for a gain material slab between the optical pump method and its homogeneous pumping rate simplification and results show that this simplification can be valid for a thin gain slab.

A self-consistent model incorporating the gain into dispersive metamaterial nanostructure was proposed and numerically solved. We numerically show that the losses of dispersive metamaterials can be compensated by gain by investigating the transmission, reflection and absorption data as well as the retrieved effective parameters. There is a relatively wide range of input signal amplitudes where the metamaterial-gain system behaves linearly. When the amplitudes get higher, the system becomes nonlinear due to the nonlinearity of the gain material itself. It's necessary to have self-consistent calculations to determine the signal range where we can expect linear response. Further, if we have strong signals so that we are in nonlinear regime or we want to study lasing, self-consistent calculation is needed. As examples, two SRR systems with different gain inclusions were studied. We have demonstrated that the magnetic losses of the SRRs can be easily compensated by the electric gain. The pumping rate needed to compensate the losses is much smaller than the bulk gain material. The losses of the SRR surrounded by gain can be easier to be compensated than the SRR with gain in the gap only due to more coupling with the gain. Provided that the pumping rate is high enough, the metamaterial nanostructures can have lasing.

\ack
Work at Ames Laboratory was supported by the Department of Energy (Basic Energy Sciences) under Contract No.~DE-AC02-07CH11358. This work was partially supported by the Office of Naval Research (Award No.~N00014-07-1-0359), by the Laboratory Directed Research and Development program at Sandia National Laboratories and BioTechnology group from AFRL/RXB.

\end{document}